\documentclass[lettersize,journal]{IEEEtran}
\usepackage{amsmath,amsfonts}
\usepackage{algorithmic}
\usepackage{algorithm}
\usepackage{array}
\usepackage[caption=false,font=normalsize,labelfont=sf,textfont=sf]{subfig}
\usepackage{textcomp}
\usepackage{stfloats}
\usepackage{url}
\usepackage{verbatim}
\usepackage{graphicx}
\usepackage{cite}
\usepackage{booktabs}
\usepackage{multicol}
\usepackage{graphicx}
\usepackage[export]{adjustbox}
\usepackage{array}
\usepackage{subcaption}
\usepackage{xcolor}
\newcolumntype{P}[1]{>{\centering\arraybackslash}p{#1}}

\usepackage{helvet} 

\usepackage{setspace} 

\usepackage{tikz}
\usetikzlibrary{shapes,arrows,positioning,calc}

\usepackage{afterpage}

\tikzset{
  block/.style = {rectangle, draw, fill=white, text centered, rounded corners, minimum height=2em, minimum width=1em}, 
  block2/.style = {rectangle, draw, fill=white, text centered, rounded corners, minimum height=2em, minimum width=1em},
  input/.style = {coordinate}, 
  output/.style = {coordinate},
  sum/.style = {draw, circle, inner sep=0pt, minimum size=0.5cm}, 
  sum2/.style = {draw, circle, inner sep=0pt, minimum size=0.5cm},
  arrow/.style = {->, thick, >=latex} 
}

\usepackage{hyperref}

\usepackage{bm}

\begin{document}

\bstctlcite{IEEEexample:BSTcontrol} 

\title{Odor Communication with Green Leaf Volatiles for Stress Signalling in the Internet of Plants}

\author{Fatih~Merdan,
       Ozgur B.~Akan,~\IEEEmembership{Fellow,~IEEE}

\thanks{Fatih Merdan and O.B. Akan are with the Center for neXt-generation Communications (CXC), Department of of Electrical and Electronics Engineering, Koç University, Istanbul 34450, Türkiye (e-mail:  fmerdan25@ku.edu.tr, akan@ku.edu.tr).}
\thanks{O. B. Akan is also with the Internet of Everything (IoE) Group,
Department of Engineering, University of Cambridge, Cambridge CB3 0FA,
U.K. (e-mail: oba21@cam.ac.uk).}}

\markboth{}%
{}

\maketitle

\begin{abstract}

This paper develops an end-to-end odor communication model for stress signaling between plants using Green Leaf Volatiles (GLV). A damaged transmitter plant emits (Z)-3-hexenal, (Z)-3-hexenol, and (Z)-3-hexenyl acetate, which propagate through a time-varying diffusion–advection channel and undergo multiplicative loss at the receiver. The sink plant is modeled with a biochemical receiver network that converts the received GLVs into the defensive metabolite (Z)-3-hexenyl $\beta$-vicianoside, and an alarm decision is defined based on its concentration level. Numerical results show that (Z)-3-hexenol is the primary driver of the system and that plant perception generally operates in a non-linear region. These findings provide a framework for understanding the evolution of plant-plant communication and for developing next-generation precision farming technologies.

\end{abstract}

\begin{IEEEkeywords}
Molecular communication, mathematical models, biological processes, plant communication, volatile organic compounds.
\end{IEEEkeywords}

\section{Introduction}
\IEEEPARstart{P}{lants} are dynamic organisms that actively interact with their environment through a set of chemical signals \cite{Defence_Priming_in_Arabidopsis_a_MetaAnalysis}. To survive various stresses, such as herbivore attacks or extreme temperatures, plants have evolved the ability to release Volatile Organic Compounds (VOCs) \cite{Herbivorous_Caterpillars_and_the_Green_Leaf_Volatile_GLV_Quandary}. Among these, Green Leaf Volatiles (GLVs) play a critical role in the rapid stress response. When a plant is physically damaged, these molecules are synthesized and released, serving as an airborne "alarm" that neighboring unstressed plants can sense to trigger preemptive defense mechanisms \cite{How_Do_Plants_Sense_Volatiles_Sent_by_Other_Plants}, \cite{PlantPlant_Communication_Is_There_a_Role_for_Volatile_DamageAssociated_Molecular_Patterns}. Characterization of these biological processes through the lens of Information and Communication Technology has emerged as a promising frontier \cite{Information_and_Communication_Theoretical_Foundations_of_the_Internet_of_Plants_Principles_Challenges_and_Future_Directions}. This interdisciplinary approach allows for the quantification of information flow within ecosystems and provides a framework for designing bio-compatible communication systems \cite{Sustainable_and_Precision_Agriculture_with_the_Internet_of_Everything_IoE}. From a communication-theoretic perspective, this interplant interaction can be modeled as a specialized Odor Communication framework \cite{Odor_Based_Molecular_Communications_State_of_the_Art_Vision_Challenges_and_Frontier_Directions}. In this framework, the stressed plant acts as a transmitter. The atmosphere serves as a time-varying physical channel where signal propagation is governed by the laws of advection and diffusion. Finally, the neighboring unstressed plant functions as a biological receiver that must decode these airborne chemical cues through complex internal metabolic pathways to mount an appropriate physiological response.

Early foundational work in this area established an end-to-end model for long-range pheromone communication between plants, characterizing the dispersion of molecules under the laws of turbulent diffusion \cite{An_EndtoEnd_Model_of_Plant_Pheromone_Channel_for_Long_Range_Molecular_Communication}. More recently, \cite{EndtoEnd_Mathematical_Modeling_of_Stress_Communication_Between_Plants} explored the context of stress communication, using continuous gene regulation to approximate Biological Volatile Organic Compound (BVOC) emissions and investigate modulation techniques. Furthermore, \cite{Modeling_and_Analysis_of_VOCbased_Interplant_Molecular_Communication_Channel} evaluates end-to-end behavior via attenuation and delay, emphasizing low-pass characteristics and distance/wind sensitivity. Despite these significant advancements, existing frameworks often simplify the receiver side of the link, frequently treating the sink plant as a basic detector or a generic absorption point. Furthermore, many models rely on static or simplified channel assumptions that do not fully capture airborne particle communication. In this paper, three primary contributions are presented. First, a rigorous biochemical receiver model is developed that is rooted in recent biological experimental data. Specifically, the internal metabolic transformation of Green Leaf Volatiles comprising (Z)-3-hexenal, (Z)-3-hexenol, and (Z)-3-hexenyl acetate into the defensive glycoside (Z)-3-hexenyl $\beta$-vicianoside is characterized. Through modeling of these metabolic pathways, (Z)-3-hexenol is identified as the main driver of the defensive state, providing a biological basis for signal reception. Second, a completely characterized time-varying channel impulse response for a diffusion-advection channel is utilized for the first time for odor communication. Finally, the analysis is extended beyond a point-to-point link to inspect the propagation of the alarm signal across a plant population for the first time.

The remainder of this paper is organized as follows. In Section II, the odor pathway in plants is discussed. Section III describes the mathematical framework for the perception of odors by plants. In Section IV, the simulation results are given and discussed. Concluding remarks are given in Section V.

\section{Odor Signalling Pathway in Plants}

\begin{figure*}[!t]
    \centering
    \includegraphics[width=\textwidth]{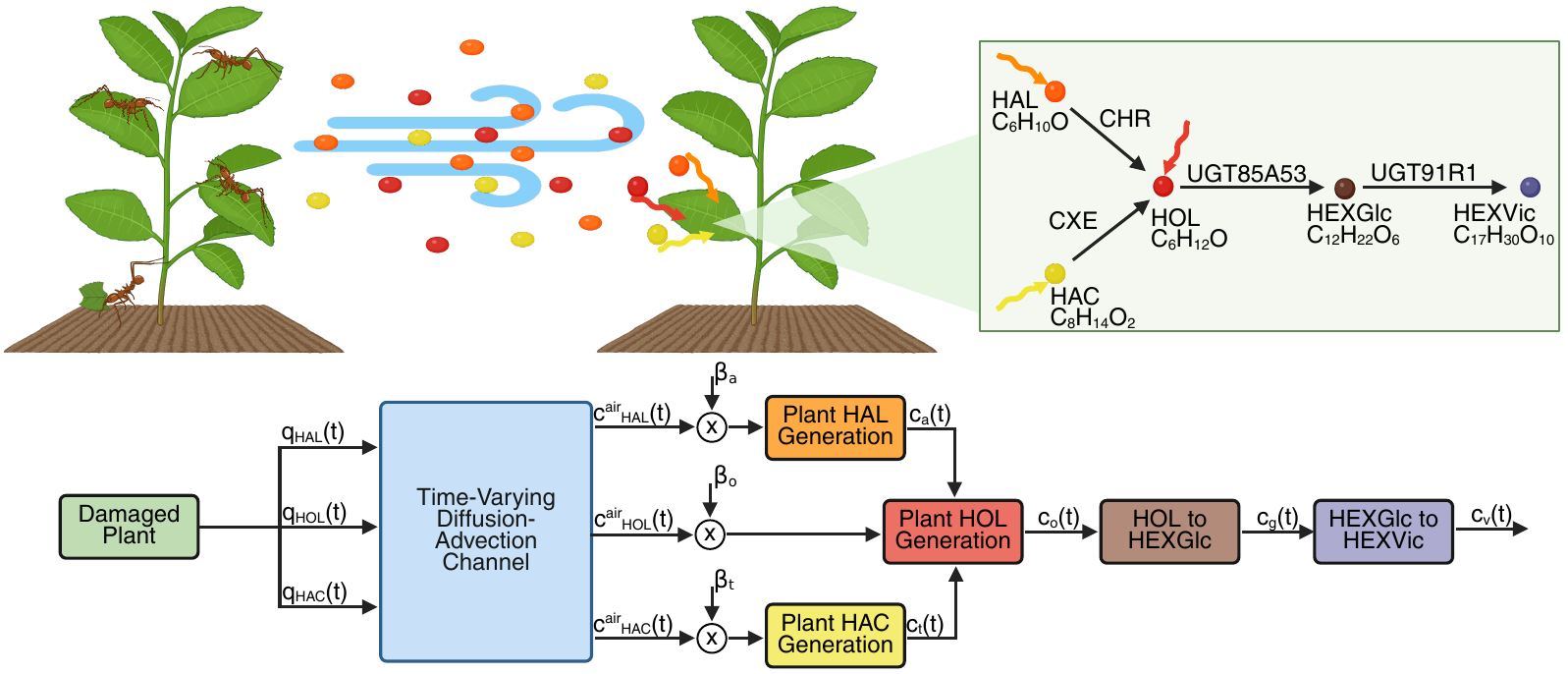}
    \caption{End-to-end odor communication model used in this work. A damaged plant emits three GLVs (HAL, HOL, and HAC) that propagate through a time-varying diffusion–advection channel. The received air concentrations are subject to multiplicative loss and drive a receiver-side biochemical network (HAL/HAC conversion to HOL and downstream conversion to HEXVic), whose output $c_v(t)$ is used for the alarm decision \cite{BioRender_Merdan_2026_3}.}
    \label{Plant_odor_Comm_system}
\end{figure*}

Plants face many different stresses throughout their life cycles. These can mainly be categorized as biotic and abiotic stresses. Biotic stresses can be caused by insects, bacteria, viruses,
fungi, nematodes, arachnids, and weeds. Abiotic stresses occur under drought, extreme temperatures, and salinity \cite{StressInduced_Volatile_Emissions_and_Signalling_in_InterPlant_Communication}. As a defense mechanism, plants release Volatile Organic Compounds (VOC) under these stress factors \cite{EndtoEnd_Mathematical_Modeling_of_Stress_Communication_Between_Plants}. Stress-induced volatiles are categorized by metabolic origin into four classes: fatty acid degradation products, such as herbivore-induced (Z)-3-hexenol; phenylpropanoids, including cold or drought-induced methyl salicylate (MeSA); biotic and abiotic stress-triggered monoterpenes such as linalool; and sesquiterpenes, such as nerolidol \cite{Volatile_CompoundMediated_PlantPlant_Interactions_under_Stress_with_the_Tea_Plant_as_a_Model}. Green Leaf Volatiles (GLVs) are fatty acid derived Volatile Organic Compounds consisting of six-carbon aldehydes, alcohols, and esters \cite{Green_Leaf_VolatilesThe_Forefront_of_Plant_Responses_Against_Biotic_Attack}. When the leaves of a plant are physically damaged, GLVs are created on site. Plants can generate substantial amounts of GLVs in seconds to minutes after tissue damage, in some cases reaching nearly 100 \(\mu g\) per gram of fresh weight \cite{Green_Leaf_Volatiles_A_New_Player_in_the_Protection_against_Abiotic_Stresses}. They play an important role in the direct defense of plants at wound sites and in indirect plant defense where plants can use GLVs to attract herbivore predators and establish stress communication between neighbouring plants \cite{Identification_of_a_Hexenal_Reductase_That_Modulates_the_Composition_of_Green_Leaf_Volatiles}, \cite{Complex_Odor_from_Plants_under_Attack_Herbivores_Enemies_React_to_the_Whole_Not_Its_Parts}.

In this study, GLV transduction is modelled between stressed and unstressed plants. The stressed plant is considered the source of GLVs, and the neighbouring plants are the sinks. Specifically, herbivores feed on the source plant, and with each bite, a new plume of GLVs is released into the air. Green leaf volatile biosynthesis initiates with the formation of (Z)-3-hexenal, which is partially reduced to (Z)-3-hexenol and subsequently acetylated by a BAHD acyltransferase to yield (Z)-3-hexenyl acetate \cite{Uptake_and_Conversion_of_Volatile_Compounds_in_PlantPlant_Communication}, \cite{Green_Leaf_VolatilesThe_Forefront_of_Plant_Responses_Against_Biotic_Attack}. (Z)-3-hexenal is an aldehyde, (Z)-3-hexenol is an alcohol, and (Z)-3-hexenyl acetate is an ester, and it is possible to transform these into one another.

After the GLVs are released into the atmosphere, they move through a diffusion-advection channel before reaching a GLV sink plant. Particle transport in flowing media arises from both diffusion and advection; however, advection dominates at the macroscale. In nature, the open air is not controlled, and the variations in wind velocity over time are modelled using stochastic signals \cite{Channel_Modeling_for_Diffusive_Molecular_CommunicationA_Tutorial_Review}.

When the released GLVs reach a GLV sink plant, they are taken by the leaves of the plant. The general route of gas exchange in plants is through the stomata, which transport $H_2O$, $CO_2$, and $O_2$. There is also evidence that GLVs are absorbed through the stomata. Adsorption on the cuticle or outer surface of the leaf is another pathway for the intake of GLV by plants \cite{Uptake_and_Conversion_of_Volatile_Compounds_in_PlantPlant_Communication}. Continuous uptake through the stomata followed by metabolic conversion is the most effective VOC intake pathway, as adsorption is considerably slower than stomatal absorption \cite{An_Absorption_Model_of_Volatile_Organic_Compound_by_Plant_Leaf_The_Most_Influential_Site_in_the_Absorption_Pathway}. This also means that the ability of plants to perceive odor molecules is highly restricted when the stomata of plants are closed, such as in the dark or under drought stress \cite{Plant_Communication_across_Different_Environmental_Contexts_Suggests_a_Role_for_Stomata_in_Volatile_Perception}. When the stomata are open, the GLVs first cross the turbulent and leaf boundary layers, enter the stomata, and diffuse into intercellular air spaces. They then undergo air–liquid partitioning, a process governed by Henry’s law constant. In the end, they reach a metabolic site in the leaf. Here, both (Z)-3-hexenal and (Z)-3-hexenyl acetate are transformed into (Z)-3-hexenol. At the same time, (Z)-3-hexenol is processed into (Z)-3-Hexenyl-\(\beta\)-D-glucopyranoside (HEXGlc). Finally, HexGlc is converted to (Z)-3-hexenyl \(\beta\)-vicianoside (HEXVic). It has been experimentally shown that HEXVic accumulating leaves are more resilient against herbivore attacks \cite{Intake_and_Transformation_to_a_Glycoside_of_Z3Hexenol_from_Infested_Neighbors_Reveals_a_Mode_of_Plant_Odor_Reception_and_Defense}. This confirms that odor transmission is used for stress communication in plants in nature. This paper focuses on HEXVic accumulation in GLV sink plants after GLV transmission from damaged GLV source plants. However, the effects of GLVs are not limited to HEXVic accumulation. For example, (Z)-3-hexenal exposure results in a \(Ca^{2+}_{cyt}\) increase in guard cells and subsequently in mesophyll cells that trigger the expression of biotic and abiotic stress-responsive genes in a \(Ca^{2+}\) dependent manner \cite{Green_Leaf_Volatile_Sensory_Calcium_Transduction_in_Arabidopsis}. Moreover, it is revealed that exposure to (Z)-3-hexenyl acetate induces ROS Production in Wheat \cite{Metabolomics_Reveal_Induction_of_ROS_Production_and_Glycosylation_Events_in_Wheat_Upon_Exposure_to_the_Green_Leaf_Volatile_Z3Hexenyl_Acetate}. Both \(Ca^{2+}\) and ROS are important secondary signals in plants that can activate many defense-related enzymes, and future research is needed to uncover their mechanics.

\section{Odor Perception Dynamics in Plants}

In a communication paradigm, there are three main elements: the transmitter, the receiver, and the channel. In this section, odor signalling between plants is characterized by mathematically defining these elements together with the relevant signals. The proposed model for the odor perception by plants is given in \autoref{Plant_odor_Comm_system}.

Due to the absence of a comprehensive dataset for a single species, biological parameters were aggregated from multiple plant sources to parameterize the system. Nevertheless, since all of these species exhibit odor perception, the resulting framework can be regarded as a general plant model for odor sensing. In addition, all odor molecules are assumed to be processed within a single effective cytosolic volume, and the corresponding enzyme concentrations are computed on the same basis.

\subsection{Odor Transmitter}

The transmitter plant releases (Z)-3-hexenal (HAL), (Z)-3-hexenol (HOL), and (Z)-3-hexenyl acetate (HAC) due to herbivore damage. Each bite of the leaf is represented as a bit. Bit 0 corresponds to no bite, where there is no GLV emission, and bit 1 corresponds to a bite. The release of GLVs into the air is modelled as a rectangular pulse of length 2 seconds. The emission amplitudes of each odor molecule are determined from \cite{Green_Leaf_Volatiles_in_the_AtmosphereProperties_Transformation_and_Significance} using the data from the cut leaves of the corn plant and given in \autoref{simulation_params}. The transmitted signal is described as
\begin{equation}
\label{Tx_signal}
\begin{split}
q(t) = \sum_{k=1}^{N} A_m p(t-kT_{sym}), 
\end{split}
\end{equation}

\noindent where \(A_m\) is the emission amplitude of the odor molecule, \(T_{sym}\) is the symbol period, and \(p(t)\) is a rectangular pulse of length \(T_{sym}\). Although proper bit and symbol definitions are made, plants are not able to reconstruct these bits. Eventually, the released GLVs will be understood as an alarm signal only.

\subsection{Diffusion-Advection Channel}

The movement of odor particles in the air is described with a partial differential equation as
\begin{equation}
\label{pde_diff_advec}
\begin{split}
\frac{\partial c}{\partial t}(\mathbf{x}, t)
&= D \nabla^{2} c(\mathbf{x}, t)
   - \mathbf{v}(t) \cdot \nabla c(\mathbf{x}, t)
   + S(\mathbf{x}, t),
\end{split}
\end{equation}

\noindent where $c(\mathbf{x},t)$ denotes the concentration of the particle at position \(\mathbf{x}\) and time \(t\), $D$ is the constant air diffusion coefficient, $\mathbf{v}(t)$ is the wind velocity vector, $\nabla c(\mathbf{x},t)$ and $\nabla^{2} c(\mathbf{x},t)$ represent the spatial gradient and Laplacian of the concentration, respectively, and $S(\mathbf{x},t)$ denotes the transmitter plant. The time-varying diffusion-advection channel impulse response described in \cite{Airborne_Particle_Communication_Through_Timevarying_DiffusionAdvection_Channels} is used to accurately model the movements of GLVs under a time-varying diffusion-advection channel. To do this, the transmitter plant is modelled as a point source as \(S(\mathbf{x}, t) = q(t)\,\delta(\mathbf{x} - \mathbf{x}_{0})\) where \(\mathbf{x}_{0}\) is the transmitter plant position and the initial zero condition is assumed for each odor molecule.

The air diffusion coefficients for each molecule are calculated using the Fuller method explained in \cite{The_Properties_of_Gases_and_Liquids}. The output of the channel is the concentration of the odor molecules around a receiver plant position.

\subsection{Odor Receiver}

Receivers are also modeled as points in a 3-D space, and it is assumed that the stomata of the receiver plants remain open throughout the communication process. The air concentrations of HAL, HOL, and HAC are the input of the receiver. The receiver plant first absorbs the odor molecules from the air. Then, these are metabolized into HEXVic through enzymatic reactions. The receiver decides that there is an attack nearby if the concentration of HEXVic in the leaves \(c_{v}(t)\) passes through a threshold \(c_{v0}\). Therefore, the amount of HEXVic accumulated in the leaf is used as the main output of the system.

\subsubsection{Random Multiplicative Loss}

Within the receiver leaf, not all absorbed odor molecules are utilized by the targeted biochemical pathway. A portion of these molecules may be sequestered in other plant tissues or consumed by competing metabolic reactions. Furthermore, the successful interaction between a molecule and its corresponding enzyme at the reaction site is inherently stochastic, depending on the probability of molecular collision \cite{Simulation_and_Inference_Algorithms_for_Stochastic_Biochemical_Reaction_Networks_From_Basic_Concepts_to_StateoftheArt}. In addition, it is found that certain herbivores release salivary enzymes that suppress GLV emissions from the source plant \cite{Silencing_the_Alarm_An_Insect_Salivary_Enzyme_Closes_Plant_Stomata_and_Inhibits_Volatile_Release}. These physical and biochemical uncertainties are collectively modeled at the receiver input through a random multiplicative loss factor.

To model the loss factors for HAL, HAC, and HOL, Beta-distributed random variables that are independent across time samples are employed. The specific profile of each loss factor is characterized by its mean and coefficient of variation (CV) \cite{Performance_of_Some_Estimators_of_Relative_Variability}, tailored to each molecule type. This framework serves as a biological analog to Signal-to-Noise Ratio (SNR) control in classical communication systems: the mean of the distribution scales the effective input amplitude, while the CV dictates the magnitude of stochastic fluctuations that degrade detection reliability.

\subsubsection{Odor Absorption}

Using the definitions in \cite{An_Absorption_Model_of_Volatile_Organic_Compound_by_Plant_Leaf_The_Most_Influential_Site_in_the_Absorption_Pathway}, VOC absorption by plant leaves is mathematically modelled as 
\begin{equation}
\label{VOC_absorption}
\begin{split}
(\frac{r_{ias}}{F} + \frac{1}{\left(\frac{1}{r^G} + \frac{E}{2}\right)} + \frac{10^{3} \, r_{liq} }{H P_{a}}) A =
\frac{\left(\frac{1}{r^G} - \frac{E}{2}\right)}{\left(\frac{1}{r^G} + \frac{E}{2}\right)} C_a - \frac{10^{3}}{H P_{a}} C_{ct},
\end{split}
\end{equation}

\noindent where \(r_{ias} = \frac{\Delta L_{ias} \tau}{D_A f_{ias}}\) is the intercellular airspace resistance, \(\Delta L_{ias}\) is the effective diffusion path length in intercellular air spaces, \(\tau\) is the diffusion path tortuosity, \(D_A\) is the diffusion coefficient of VOC in gas phase and \(f_{ias}\) is the fraction of the intercellular air space volume in the total leaf volume. $F = \frac{273.15}{T_k \times 22.4e-3}$ is a scaling constant where $T_k$ is the temperature in $kelvin$. \(r_G = r_s + r_b\) is the total resistance from outside air to intercellular air spaces. \(r_b = r_b^{w} (\frac{D_w}{D_A})^\frac{2}{3}\) is the resistance of the leaf boundary layer, where \(r_b^{w}\) is the resistance of the leaf boundary layer for water vapor and \(D_w\) is the diffusion coefficient of water in the gas phase. Similarly, \(r_s = r_s^{w} (\frac{D_w}{D_A})\) is the stomatal resistance and \(r_s^{w}\) is the stomatal resistance for water vapor. $E$ is the VOC specific transpiration rate, $H$ is the VOC specific Henry's constant taken from \cite{Green_Leaf_Volatiles_in_the_AtmosphereProperties_Transformation_and_Significance} and $P_a$ is the pressure in $bar$. \(r_{liq}\) is the sum of the resistances of the cell wall, plasma membrane, and cytosol. It is calculated following the description given in \cite{An_Absorption_Model_of_Volatile_Organic_Compound_by_Plant_Leaf_The_Most_Influential_Site_in_the_Absorption_Pathway} for each odor molecule. To do this, the parameters of the plant \textit{Spathiphyllum clevelandii} are used. Moreover, the Octanol/Water partition coefficients (\(K_{o/w}\)) for the odor molecules are obtained from the Chemeo database \cite{chemeo_hexenal,chemeo_hexenol,chemeo_hexenylacetate}.

In (\ref{VOC_absorption}), $A$ is the odor molecule absorption rate in \(mol \, m^{-2} s^{-1}\), \(C_a\) is the odor concentration in the atmosphere in \(ppb\) and \(C_{ct}\) is the odor concentration in cytosol in \(mol \, m^{-3}\). In this equation, \(C_a\) and \(C_{ct}\) are known, and $A$ is the output at a time instant. The output of the channel is in \(\mathrm{mol/m^3}\). To convert the unit of \(C_a\) from \(\mathrm{mol/m^3}\) to \(ppb\), the methods explained in \cite{epa_conversion} and \cite{Atmospheric_Chemistry_and_Physics_From_Air_Pollution_to_Climate_Change} are utilized. Since in the receiver, \(molar\) absorption rates are used, \(A\) is converted into \(mol \, L^{-1} s^{-1}\) as
\begin{equation}
\label{VOC_absorption_unit_conversion}
\begin{split}
A (mol \, L^{-1} s^{-1}) = A (mol \, m^{-2} s^{-1}) \times \frac{LA_{FW}}{V_{intra}},
\end{split}
\end{equation}

\noindent where \(LA_{FW}\) is the leaf area per fresh weight (\(m^2 g^{-1} \, FW\)) obtained from \cite{The_Arabidopsis_Leaf_Quantitative_Atlas_A_Cellular_and_Subcellular_Mapping_through_Unified_Data_Integration} and \(V_{intra}\) is the water volume of leaf tissue per unit fresh weight (\(L \, g^{-1} \, FW\)). Using the fact that \textit{Arabidopsis} leaves normally has 0.90 \(g \, g^{-1} \, FW\) \cite{Water_Deficit_Enhances_C_Export_to_the_Roots_in_Arabidopsis_Thaliana_Plants_with_Contribution_of_Sucrose_Transporters_in_Both_Shoot_and_Roots1OPEN}, and taking the density of water \(1 \, g/ml\), \(V_{intra}\) is calculated.

\subsubsection{HAL to HOL Conversion}
Absorbed HAL inside the leaves ($c_a(t)$) is processed into HOL. In \cite{Identification_of_a_Hexenal_Reductase_That_Modulates_the_Composition_of_Green_Leaf_Volatiles}, the enzyme encoded by At4g37980 in \textit{Arabidopsis} is shown to convert HAL into HOL, and is named as CHR. To represent this chemical reaction, Michaelis-Menten Equations are used \cite{Enzyme_Kinetics_Behavior_and_Analysis_of_Rapid_Equilibrium_and_SteadyState_Enzyme_Systems_National_Library_of_Medicine_Institution}. The rate of change from HAL to HOL is represented as 
\begin{equation}
\label{HAL_to_HOL_eqn}
\begin{split}
J_{AO}(t) = k_{cat}^{ao} E_{CHR} \frac{c_a(t)}{k_{m}^{ao} + c_a(t)},
\end{split}
\end{equation}

\noindent where \(k_{cat}^{ao}\) is the catalytic constant, \(E_{CHR}\) is the total CHR concentration in the leaf,  \(k_{m}^{ao}\) is the Michaelis constant for this reaction and \(c_a(t)\) is the total cytosolic HAL concentration in the leaf.

\subsubsection{HAC to HOL Conversion}
As reported in \cite{The_Carboxylesterase_AtCXE12_Converts_Volatile_Z3Hexenyl_Acetate_to_Z3Hexenol_in_Arabidopsis_Leaves}, the Arabidopsis enzyme AtCXE12 is shown to metabolize HAC into HOL. To model this reaction, a similar Michaelis-Menten dynamics is used as
\begin{equation}
\label{HAC_to_HOL_eqn}
\begin{split}
J_{TO}(t) = k_{cat}^{to} E_{CXE} \frac{c_t(t)}{k_{m}^{to} + c_t(t)},
\end{split}
\end{equation}

\noindent where  \(E_{CXE}\) is the total AtCXE12 concentration in the leaf,  and \(c_t(t)\) is the total cytosolic HAC concentration in the leaf.

\subsubsection{HOL to HEXGlc Conversion}

According to \cite{Glucosylation_of_Z3Hexenol_Informs_Intraspecies_Interactions_in_Plants_A_Case_Study_in_Camellia_Sinensis}, the UGT85A53 protein in \textit{Camellia sinensis} is responsible for HEXGlc accumulation from airborne HOL. A Michaelis-Menten dynamics is used as
\begin{equation}
\label{HOL_to_HexGlc_eqn}
\begin{split}
J_{OG}(t) = k_{cat}^{og} E_{85A} \frac{c_o(t)}{k_{m}^{og} + c_o(t)},
\end{split}
\end{equation}

\noindent where  \(E_{85A}\) is the total UGT85A53 concentration in the leaf,  and \(c_o(t)\) is the total cytosolic HOL concentration in the leaf.

\subsubsection{HEXGlc to HEXVic Conversion}

Evidence in \cite{Identification_of_a_Tomato_UDParabinosyltransferase_for_Airborne_Volatile_Reception} shows that the enzyme UGT91R1 mediates HEXVic accumulation in \textit{S. lycopersicum} by preferentially catalyzing the arabinosylation for HEXGlc. Similarly, this dynamic is modelled as 
\begin{equation}
\label{HexGlc_to_HexVic_eqn}
\begin{split}
J_{GV}(t) = k_{cat}^{gv} E_{91R} \frac{c_g(t)}{k_{m}^{gv} + c_g(t)},
\end{split}
\end{equation}

\noindent where  \(E_{91R}\) is the total UGT91R1 concentration in the leaf,  and \(c_g(t)\) is the total cytosolic HEXGlc concentration in the leaf. \(c_v(t)\) is used to represent the total cytosolic HEXVic concentration.

Using the method explained in \cite{Protein_Abundance_Biases_the_Amino_Acid_Composition_of_Disordered_Regions_to_Minimize_Nonfunctional_Interactions}, the total enzyme concentrations from \(ppm\) values can be found as
\begin{equation}
\label{ppm_to_total_enzyme_eqn}
\begin{split}
E_{tot} = \frac{k_e A_b 10^{15}}{N_A},
\end{split}
\end{equation}

\noindent where \(k_e \approx 3e6\) is an estimate of cellular density of protein molecules per \(femtoliter\), \(A_b\) is the abundance, and $N_A$ is the Avogadro constant. The abundance values of the enzymes are obtained from the PaxDb database \cite{paxdb}. For AtCXE12 and CHR, the Arabidopsis leaf tissue is used. For UGT85A53, the Arabidopsis gene UGT85A1, and for UGT91R1, the Arabidopsis gene UGT91A1 is used, with the abundance in the whole organism.

Using the absorption rates from (\ref{VOC_absorption_unit_conversion}) and the internal enzymatic rates from  (\ref{HAL_to_HOL_eqn}),  (\ref{HAC_to_HOL_eqn}),  (\ref{HOL_to_HexGlc_eqn}), and  (\ref{HexGlc_to_HexVic_eqn}) the following system of differential equations is constructed
\begin{equation}
\label{mass_balances}
\begin{split}
\frac{d c_a(t)}{dt} &= A_a(t) - J_{AO}(t),\\[4pt]
\frac{d c_t(t)}{dt} &= A_t(t) - J_{TO}(t),\\[4pt]
\frac{d c_o(t)}{dt} &= A_o(t) + J_{AO}(t) + J_{TO}(t) - J_{OG}(t),\\[4pt]
\frac{d c_g(t)}{dt} &= J_{OG}(t) - J_{GV}(t),\\[4pt]
\frac{d c_v(t)}{dt} &= J_{GV}(t).
\end{split}
\end{equation}

\noindent where \(A_a(t)\), \(A_o(t)\), and \(A_t(t)\) are the absorption rates of HAL, HOL, and HAC, respectively. All biological-chemical parameters are given in \autoref{Biological_params}.

\begin{table*}[!t]
  \caption{Biological-Chemical Parameters}
  \centering
  \label{Biological_params}
  \footnotesize 
  \renewcommand{\arraystretch}{1.15} 
  \setlength{\tabcolsep}{4pt}
  \begin{tabular}{
    @{}>{\raggedright\arraybackslash\itshape}p{4.2cm} 
    >{\raggedright\arraybackslash\itshape}p{0.8cm}
    >{\raggedright\arraybackslash}p{1.8cm}
    >{\raggedright\arraybackslash}p{1.6cm}
    | 
    >{\raggedright\arraybackslash\itshape}p{4.4cm} 
    >{\raggedright\arraybackslash\itshape}p{0.8cm}
    >{\raggedright\arraybackslash}p{1.2cm}
    >{\raggedright\arraybackslash}p{1.6cm}@{}}
    \toprule

    \textbf{Description} & \textbf{Symbol} & \textbf{Value} & \textbf{Unit} &
    \textbf{Description} & \textbf{Symbol} & \textbf{Value} & \textbf{Unit} \\
    \midrule

    Temperature & \(T_k\) & 298.15 & \(kelvin\) &
    Diffusion path tortuosity & \(\tau\) & 1.57 & \(m m^{-1}\) \\

    Pressure & \(P\) & 1 & \(atm\) &
    Fraction of the intercellular air space volume in the total leaf volume & \(f_{ias}\) & 0.328 & \(m^3 m^{-3}\) \\

    Molecular weight of HAL & \(M_a\) & 98.143 & \(g/mol\) &
    Henry’s constant (HAL) & \(H_{a}\) & 6.0 & \(mol \, L^{-1} atm^{-1}\) \\

    Molecular weight of HOL & \(M_o\) & 100.159 & \(g/mol\) &
    Henry’s constant (HOL) & \(H_{o}\) & 113 & \(mol \, L^{-1} atm^{-1}\) \\

    Molecular weight of HAC & \(M_t\) & 142.2 & \(g/mol\) &
    Henry’s constant (HAC) & \(H_{t}\) & 3.1 & \(mol \, L^{-1} atm^{-1}\) \\

    Molecular weight of vapor & \(M_{w}\) & 18 & \(g/mol\) &
    Octanol/Water partition coefficient (HAL) & \(K_{o/w}^{a}\) & \(10^{1.542}\) & - \\

    Diffusion coefficient of HAL & \(D_a\) & $8.0718\times10^{-6}$ & \(m^2/s\) &
    Octanol/Water partition coefficient (HOL) & \(K_{o/w}^{o}\) & \(10^{1.335}\) & - \\

    Diffusion coefficient of HOL & \(D_o\) & $7.9291\times10^{-6}$ & \(m^2/s\) &
    Octanol/Water partition coefficient (HAC) & \(K_{o/w}^{t}\) & \(10^{1.906}\) & - \\

    Diffusion coefficient of HAC & \(D_t\) & $6.7698\times10^{-6}$ & \(m^2/s\) &
    Abundance of CHR & \(A_b^{CHR}\) & 330 & \(ppm\) \\

    Diffusion coefficient of vapor & \(D_{w}\) & $2.3289\times10^{-5}$ & \(m^2/s\) &
    Abundance of AtCXE12 & \(A_b^{CXE}\) & 122 & \(ppm\) \\

    Leaf area per fresh weight & \(LA_{FW}\) & 0.0055 & \(m^2 g^{-1} \, FW\) &
    Abundance of UGT85A53 & \(A_b^{85A}\) & 13.2 & \(ppm\) \\

    water volume of leaf tissue per unit fresh weight & \(V_{intra}\) & 0.0009 & \(L \, g^{-1} \, FW\) &
    Abundance of UGT91R1 & \(A_b^{91R}\) & 0.09 & \(ppm\) \\

    Leaf boundary layer resistance for water vapor (HAL) & \(r_{b_a}^{w}\) & 2.58 & \(m^2 \, s/mol\) &
    Avogadro constant & \(N_A\) & 6.02$\times10^{23}$ & \(mol^{-1}\) \\

    Stomatal resistance for water vapor (HAL) & \(r_{s_a}^{w}\) & 21.8 & \(m^2 \, s/mol\) &
    Catalytic constant (CHR) & \(k_{cat}^{ao}\) & 13.27 & \(s^{-1}\) \\

    Leaf boundary layer resistance for water vapor (HOL) & \(r_{b_o}^{w}\) & 3.23 & \(m^2 \, s/mol\) &
    Catalytic constant (CXE) & \(k_{cat}^{to}\) & 3.78 & \(s^{-1}\) \\

    Stomatal resistance for water vapor (HOL) & \(r_{s_o}^{w}\) & 26.5 & \(m^2 \, s/mol\) &
    Catalytic constant (85A) & \(k_{cat}^{og}\) & 0.35 & \(s^{-1}\) \\

    Leaf boundary layer resistance for water vapor (HAC) & \(r_{b_t}^{w}\) & 2.47 & \(m^2 \, s/mol\) &
    Catalytic constant (91R) & \(k_{cat}^{gv}\) & 0.33 & \(s^{-1}\) \\

    Stomatal resistance for water vapor (HAC) & \(r_{s_t}^{w}\) & 16.1 & \(m^2 \, s/mol\) &
    Michaelis constant (CHR) & \(k_{m}^{ao}\) & 32.7 & \(\mu M\) \\

    HAL transpiration rate & \(E_{a}\) & $6\times10^{-4}$ & \(mol \, m^{-2} s^{-1}\) &
    Michaelis constant (CXE) & \(k_{m}^{to}\) & 5940 & \(\mu M\) \\

    HOL transpiration rate & \(E_{o}\) & $5.4\times10^{-4}$ & \(mol \, m^{-2} s^{-1}\) &
    Michaelis constant (85A) & \(k_{m}^{og}\) & 18.92 & \(\mu M\) \\

    HAC transpiration rate & \(E_{t}\) & $4.5\times10^{-4}$ & \(mol \, m^{-2} s^{-1}\) &
    Michaelis constant (91R) & \(k_{m}^{gv}\) & 5.9 & \(\mu M\) \\

    Effective diffusion path length in intercellular air spaces & \(\Delta L_{ias}\) & $6.38\times10^{-5}$ & \(m\) &
    & & & \\ 

    \bottomrule
  \end{tabular}
\end{table*}

\section{Simulation Results}

In this Section, numerical analysis of the given topology is carried out. First, the receiver characteristics are inspected in detail. Then, the end-to-end communication system is inspected between one transmitter and one receiver plant. Lastly, the distribution of the alarm signal is analyzed across a comprehensive set of receiver coordinates to characterize how the warning message is perceived by the surrounding plant population under different wind conditions. For multi-receiver scenarios, the channel is simulated in Python using an NVIDIA GeForce GTX 1080 Ti GPU, and the channel output is saved. Subsequent analysis is conducted using MATLAB R2023b. The parameters used in the simulations are given in \autoref{simulation_params}.

\subsection{Receiver Characteristics}

The odor receiver presented in this work is designed to study the initial stimulation phase of the receiver plant. Because the model does not yet include a 'reset' or recovery mechanism, it continuously accumulates incoming signals. This receiver can be taken as a linear time invariant (LTI) system if for each enzymatic reaction $k_m \gg c(t)$ is satisfied. It can be observed that, as long as the receiver keeps getting an input, it will leave the linear operating region because the internal concentrations will increase as time progresses. To understand the linearity limits of this system with respect to its $3$ inputs, a practical method is introduced. Specifically, if for that simulation duration, the time fraction for which $r > 0.1$ is less than or equal to $0.02$, the system is treated as operating in the linear regime, where 
\begin{equation}
\label{lineartiy_rule}
\begin{split}
r = max( \frac{c_a(t)}{k_{m}^{ao}}, \frac{c_t(t)}{k_{m}^{to}}, \frac{c_o(t)}{k_{m}^{og}}, \frac{c_g(t)}{k_{m}^{gv}}).
\end{split}
\end{equation}

\begin{figure*}[!t]
  \centering

  \subfloat[]{
    \includegraphics[width=0.21\linewidth]{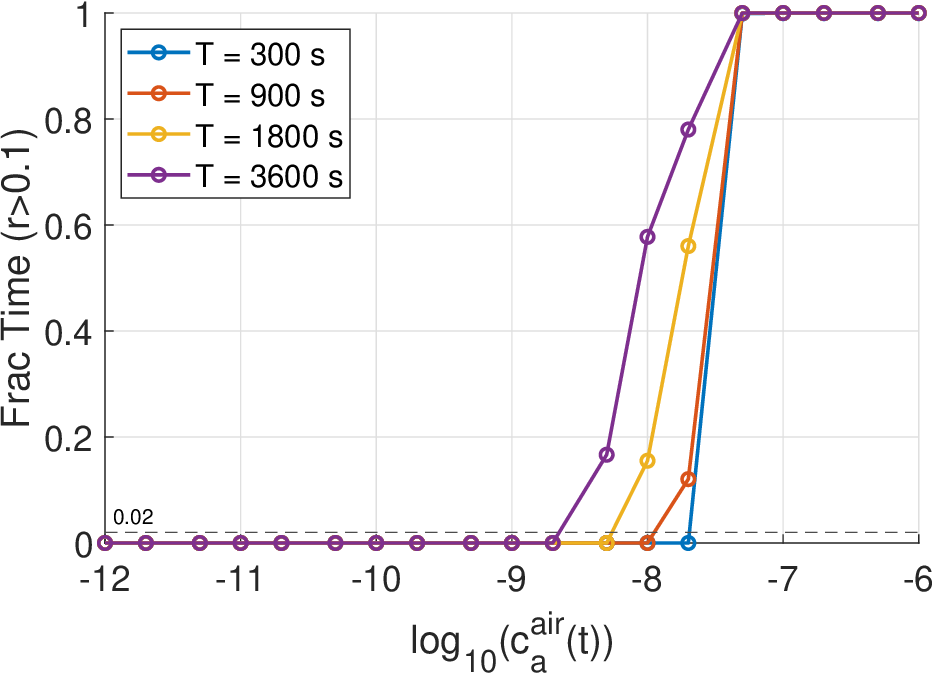}
    \label{LinearityPilot_SingleVOC_HAL_mode_1}}
  \hspace{0.01\linewidth}
  \subfloat[]{
    \includegraphics[width=0.21\linewidth]{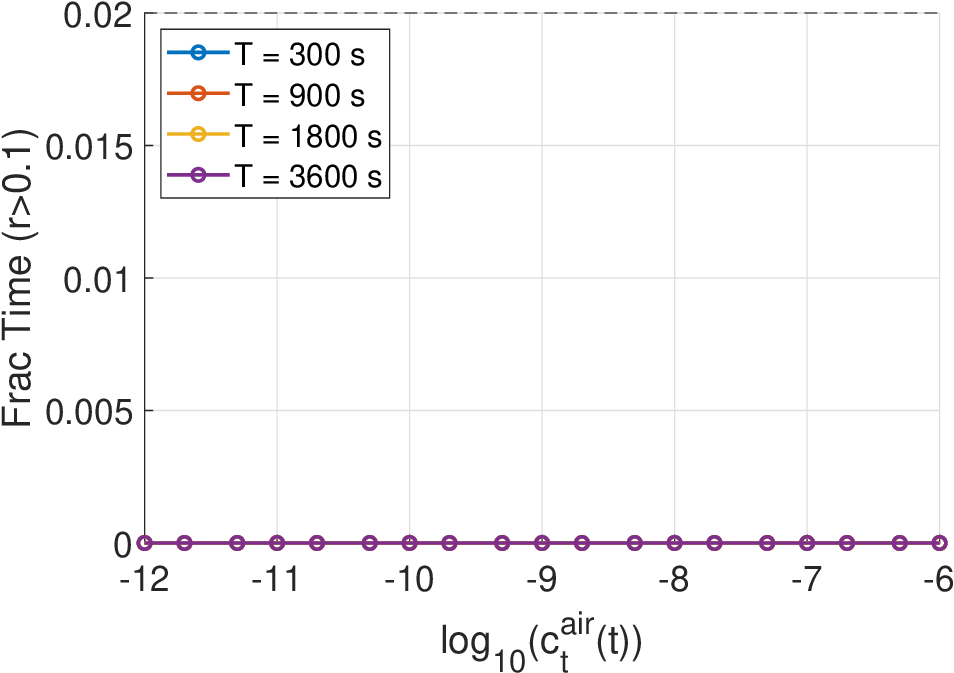}
    \label{LinearityPilot_SingleVOC_HAC_mode_1}}
  \hspace{0.01\linewidth}
  \subfloat[]{
    \includegraphics[width=0.21\linewidth]{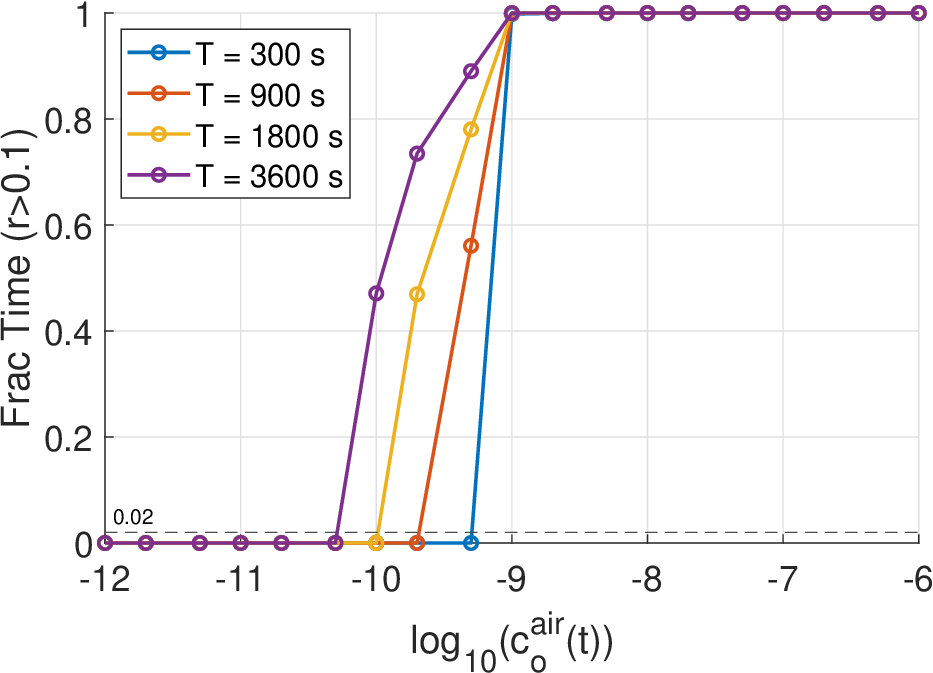}
    \label{LinearityPilot_SingleVOC_HOL_mode_1}}
  \hspace{0.01\linewidth}
  \subfloat[]{
    \includegraphics[width=0.21\linewidth]{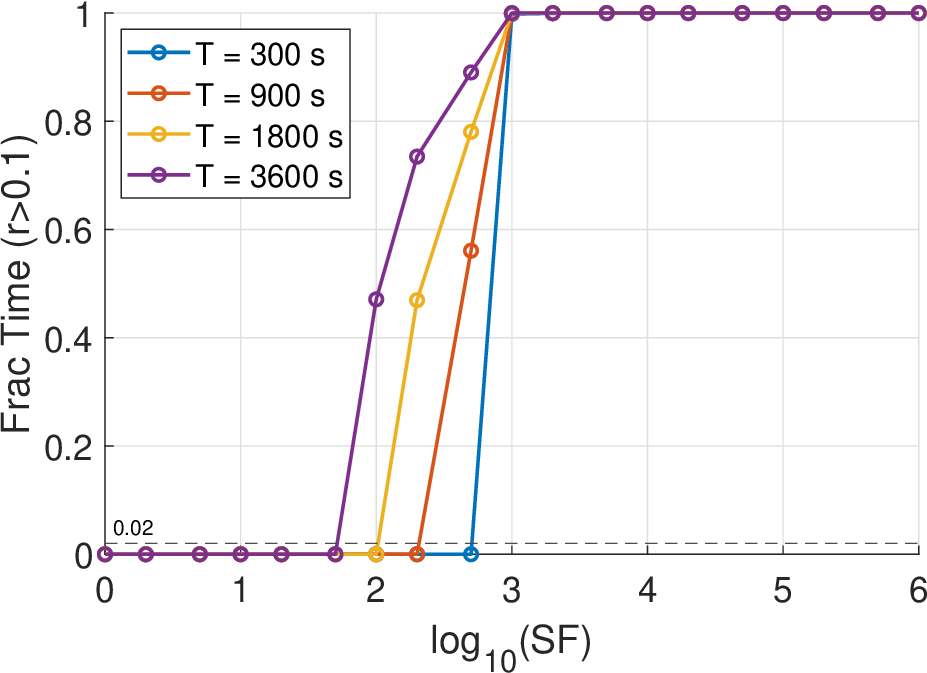}
    \label{LinearityPilot_AllVOC_mode_1}}


  \subfloat[]{
    \includegraphics[width=0.21\linewidth]{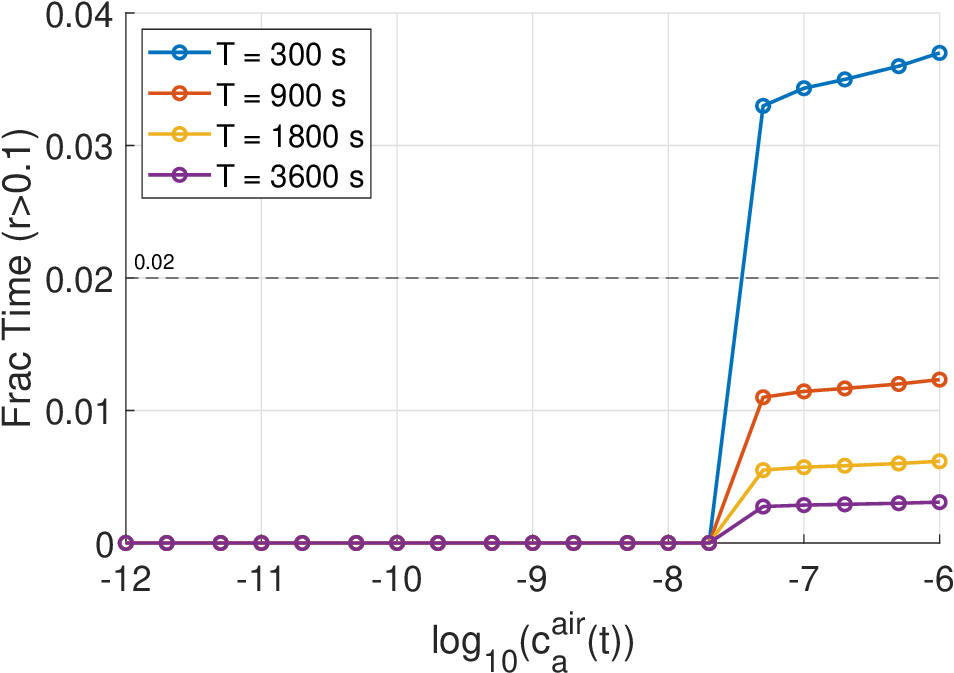}
    \label{LinearityPilot_SingleVOC_HAL_mode_2}}
  \hspace{0.01\linewidth}
  \subfloat[]{
    \includegraphics[width=0.21\linewidth]{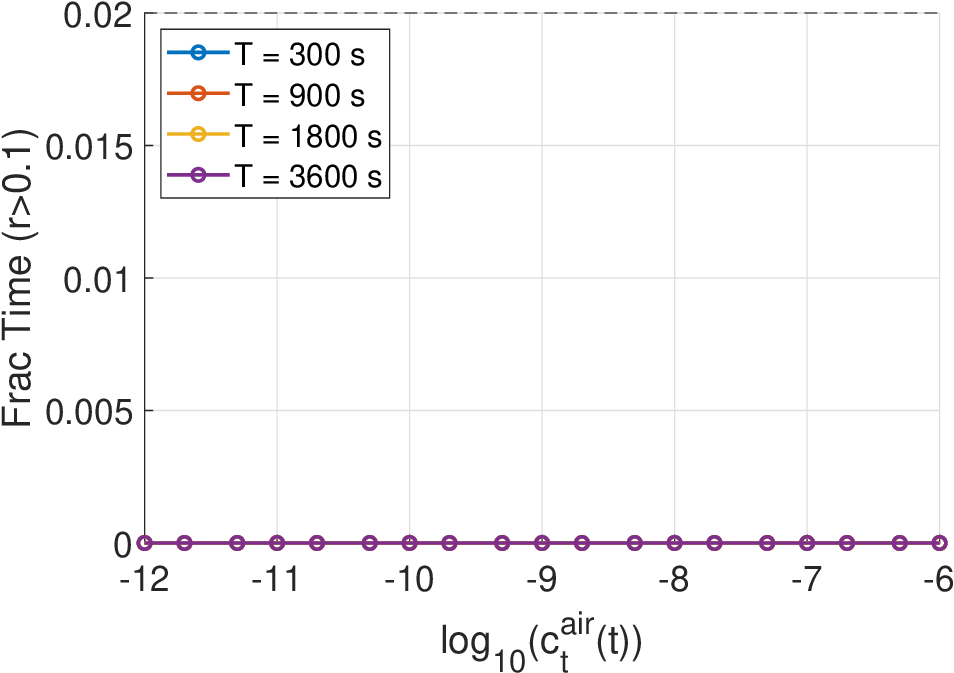}
    \label{LinearityPilot_SingleVOC_HAC_mode_2}}
  \hspace{0.01\linewidth}
  \subfloat[]{
    \includegraphics[width=0.21\linewidth]{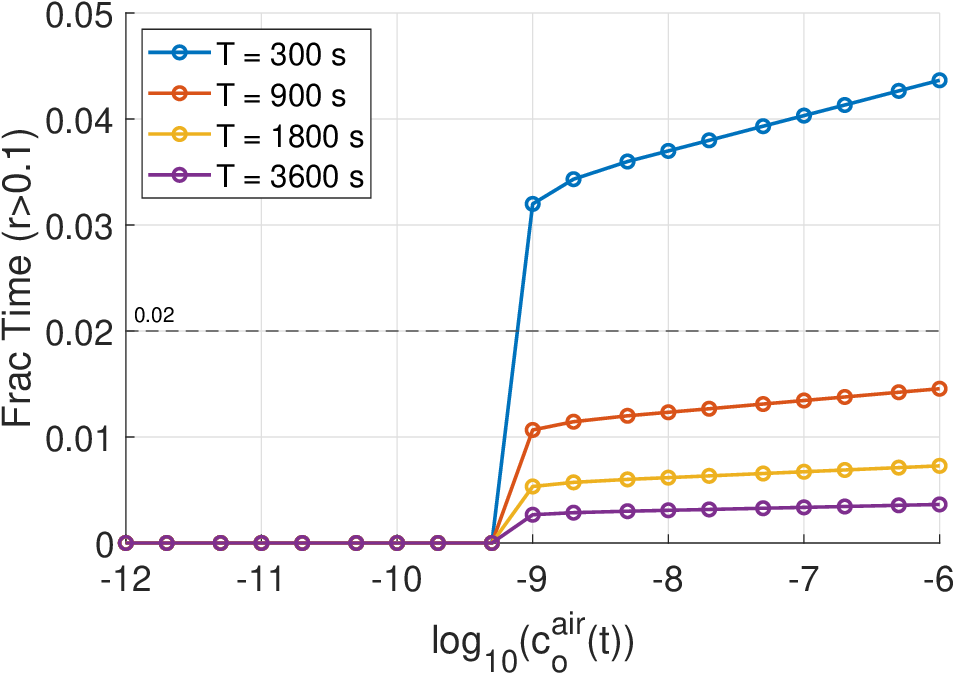}
    \label{LinearityPilot_SingleVOC_HOL_mode_2}}
  \hspace{0.01\linewidth}
  \subfloat[]{
    \includegraphics[width=0.21\linewidth]{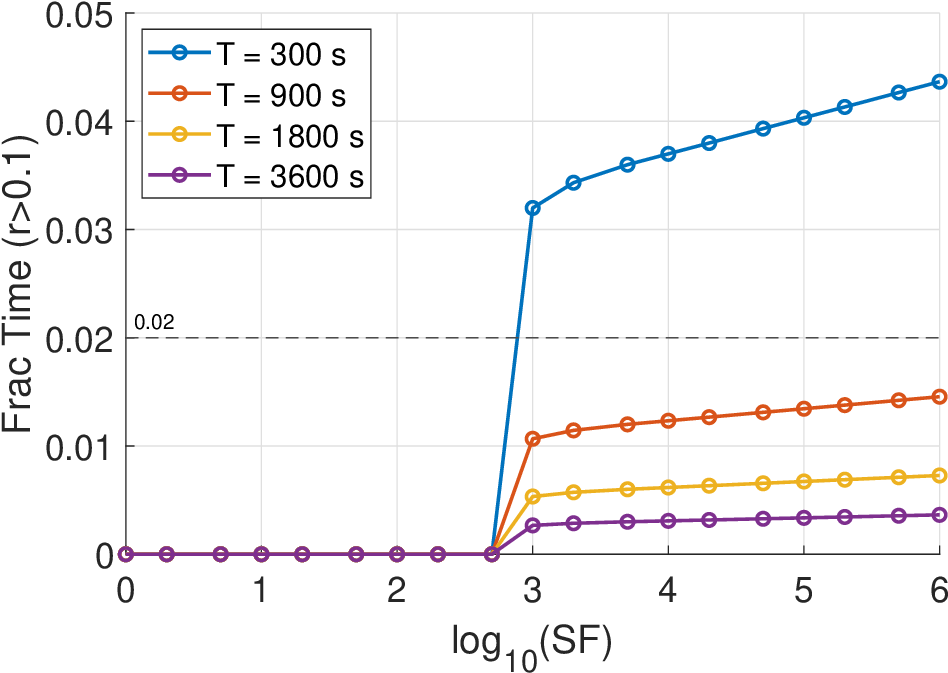}
    \label{LinearityPilot_AllVOC_mode_2}}


  \subfloat[]{
    \includegraphics[width=0.21\linewidth]{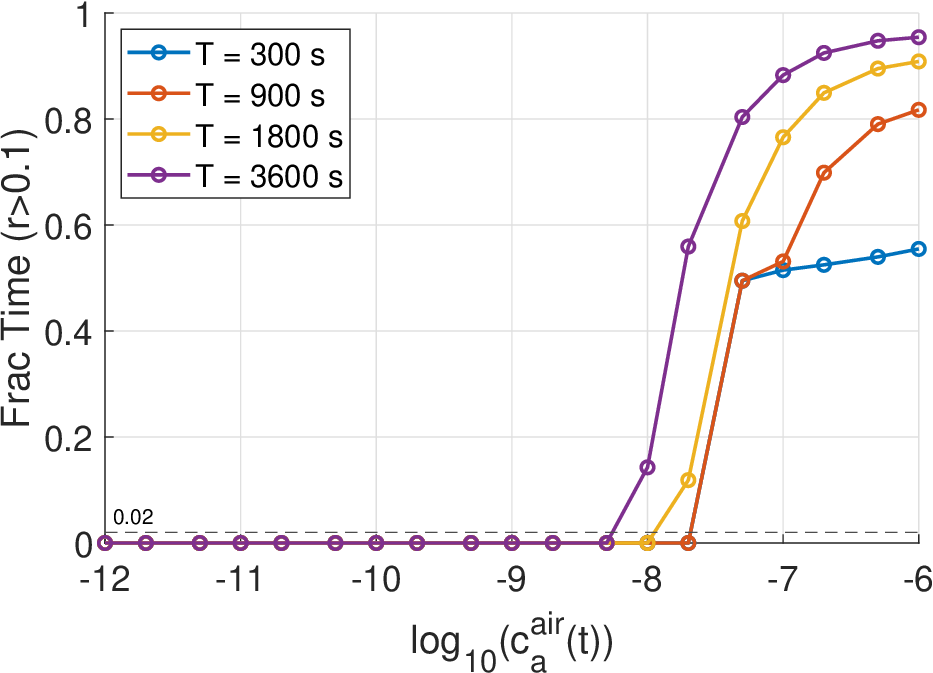}
    \label{LinearityPilot_SingleVOC_HAL_mode_3}}
  \hspace{0.01\linewidth}
  \subfloat[]{
    \includegraphics[width=0.21\linewidth]{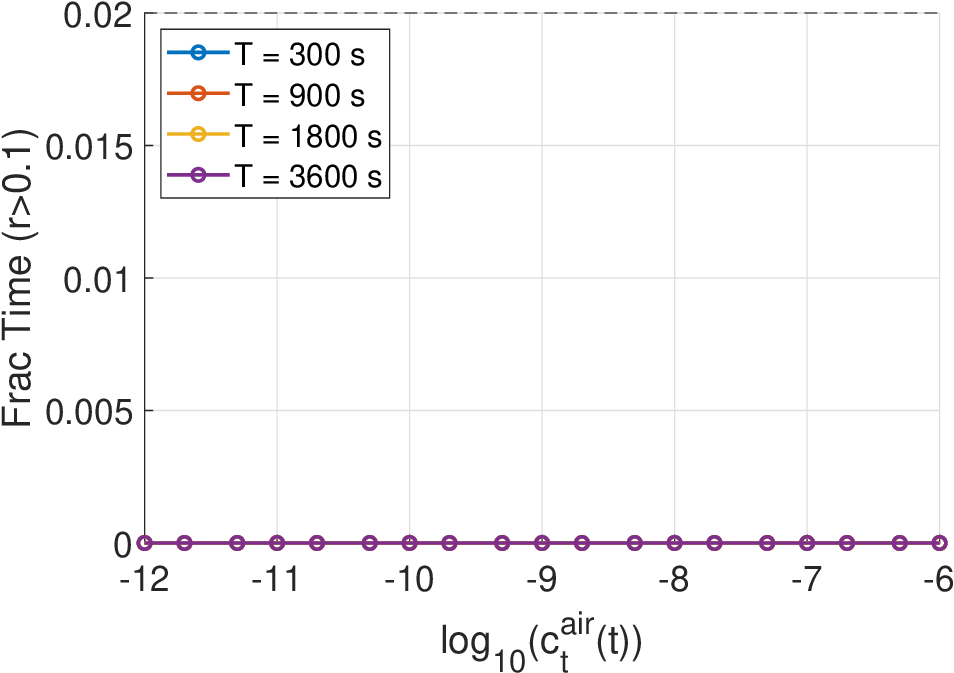}
    \label{LinearityPilot_SingleVOC_HAC_mode_3}}
  \hspace{0.01\linewidth}
  \subfloat[]{
    \includegraphics[width=0.21\linewidth]{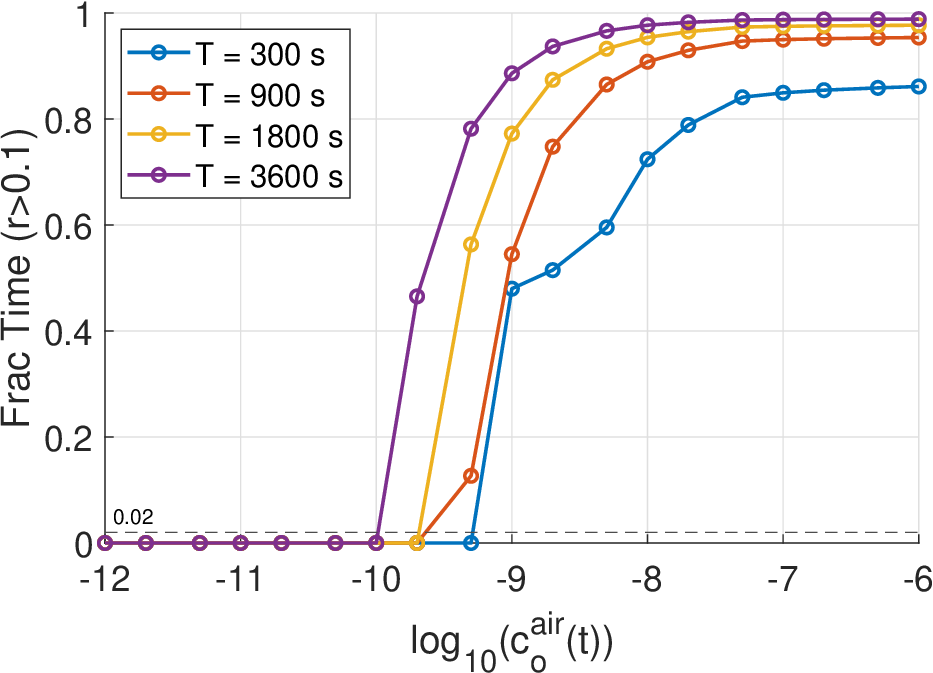}
    \label{LinearityPilot_SingleVOC_HOL_mode_3}}
  \hspace{0.01\linewidth}
  \subfloat[]{
    \includegraphics[width=0.21\linewidth]{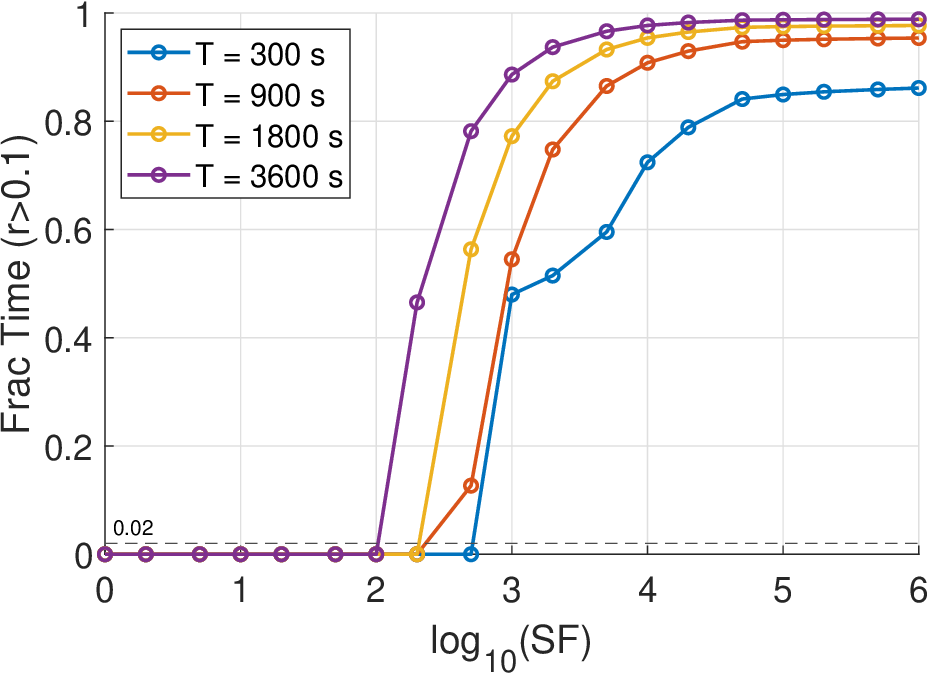}
    \label{LinearityPilot_AllVOC_mode_3}}

    \caption{Receiver nonlinearity results across various simulation times, input signal configurations, and scaling scenarios. For \textit{(i)} constant inputs \protect\subref{LinearityPilot_SingleVOC_HAL_mode_1} scales HAL only,  \protect\subref{LinearityPilot_SingleVOC_HAC_mode_1} scales HAC only, \protect\subref{LinearityPilot_SingleVOC_HOL_mode_1} scales HOL only, and \protect\subref{LinearityPilot_AllVOC_mode_1} scales all three input amplitudes together, and the horizontal axis shows the common scaling factor (SF). For \textit{(ii)} single initial pulses, \protect\subref{LinearityPilot_SingleVOC_HAL_mode_2} scales HAL only, \protect\subref{LinearityPilot_SingleVOC_HAC_mode_2} scales HAC only, \protect\subref{LinearityPilot_SingleVOC_HOL_mode_2} scales HOL only, and \protect\subref{LinearityPilot_AllVOC_mode_2} scales all three input amplitudes together. For \textit{(iii)} periodic on-off cycling, \protect\subref{LinearityPilot_SingleVOC_HAL_mode_3} scales HAL only, \protect\subref{LinearityPilot_SingleVOC_HAC_mode_3} scales HAC only, \protect\subref{LinearityPilot_SingleVOC_HOL_mode_3} scales HOL only, and
    \protect\subref{LinearityPilot_AllVOC_mode_3} scales all three input amplitudes together. The horizontal dashed lines indicate the $0.02$ threshold used to define the LTI operating region. }
  \label{LinearityPilot_grid_modes_3x4}
\end{figure*}

The linearity condition is evaluated across various simulation durations and input signal profiles. In each configuration, the amplitude of one input signal is increased while the other two are held constant to examine the linearity; subsequently, all three input amplitudes are simultaneously increased to observe the collective behavior. For all cases, the initial air concentrations of each odor molecule are set to $1\times10^{-12} \, mol/m^3$. Three input signal configurations are evaluated: \textit{(i)} a constant input, \textit{(ii)} a single initial pulse ($T_{on} = 10 s$), and \textit{(iii)} periodic on-off cycling ($T_{on} = T_{off} = 10 s$). The results are presented in \autoref{LinearityPilot_grid_modes_3x4}. As expected, for both constant and periodic on-off cycling input profiles, the minimum input amplitude required to exceed the linearity threshold decreases as simulation duration increases. In the case of a single initial pulse, the specific amplitude at which linearity breaks can be identified for each odor molecule. Notably, the system exhibits a return to the linear regime as time progresses in this configuration. This recovery occurs because the concentrations of molecules responsible for breaking linearity decrease over time as they are converted into downstream metabolites like HEXGlc. Across all configurations, the results for HOL scaling and simultaneous scaling are nearly identical, implying that HOL is the primary limiter of the linear operating region. In contrast, HAC remains within the linear regime across all tested scenarios and durations.

\begin{figure*}[!t]
  \centering

  \subfloat[]{
    \includegraphics[width=0.3\linewidth]{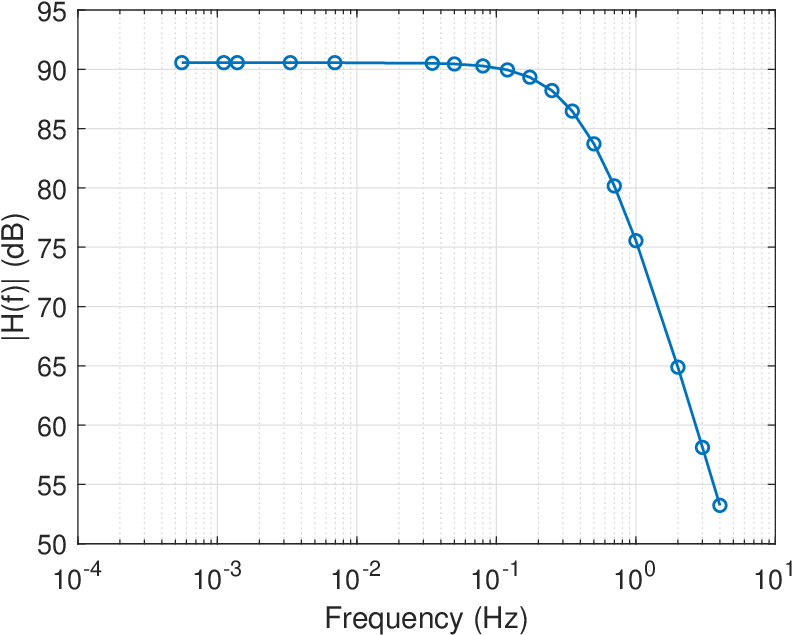}
    \label{freq_response_input_hal_to_HOL_magnitude}}
  \hspace{0.015\linewidth}
  \subfloat[]{
    \includegraphics[width=0.3\linewidth]{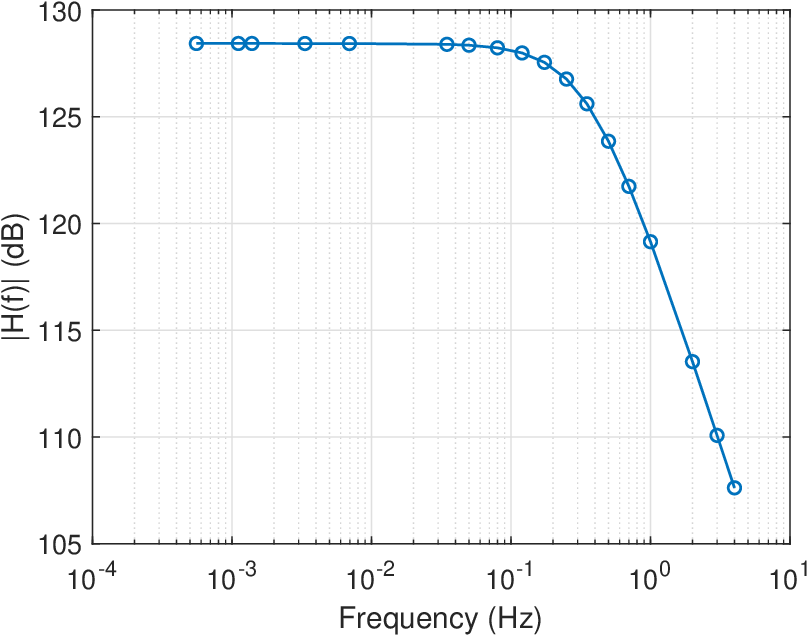}
    \label{freq_response_input_hol_to_HOL_magnitude}}
  \hspace{0.015\linewidth}
  \subfloat[]{
    \includegraphics[width=0.3\linewidth]{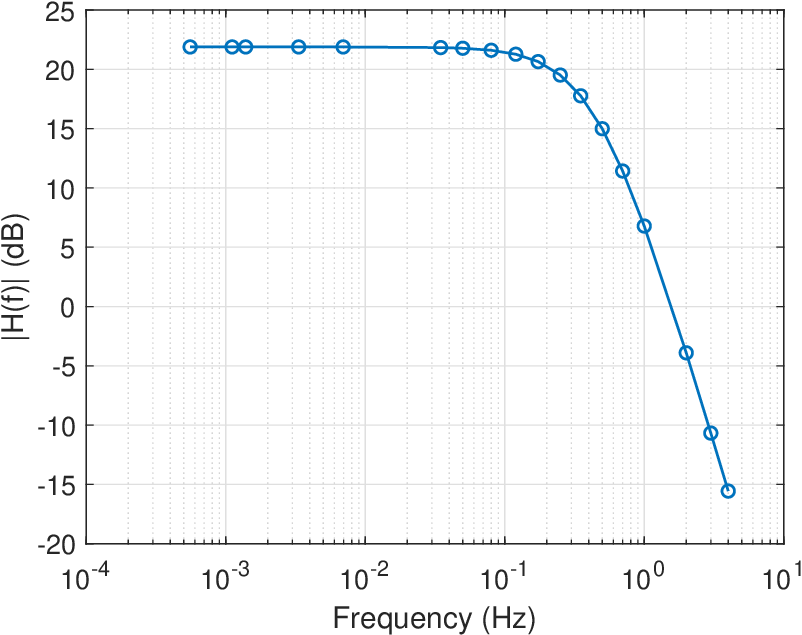}
    \label{freq_response_input_hac_to_HOL_magnitude}}


  \subfloat[]{
    \includegraphics[width=0.3\linewidth]{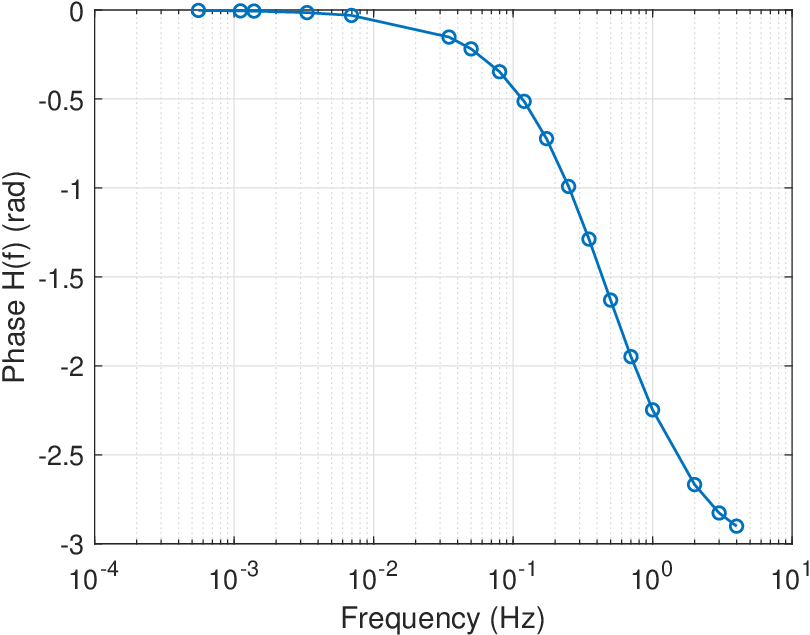}
    \label{freq_response_input_hal_to_HOL_phase}}
  \hspace{0.015\linewidth}
  \subfloat[]{
    \includegraphics[width=0.3\linewidth]{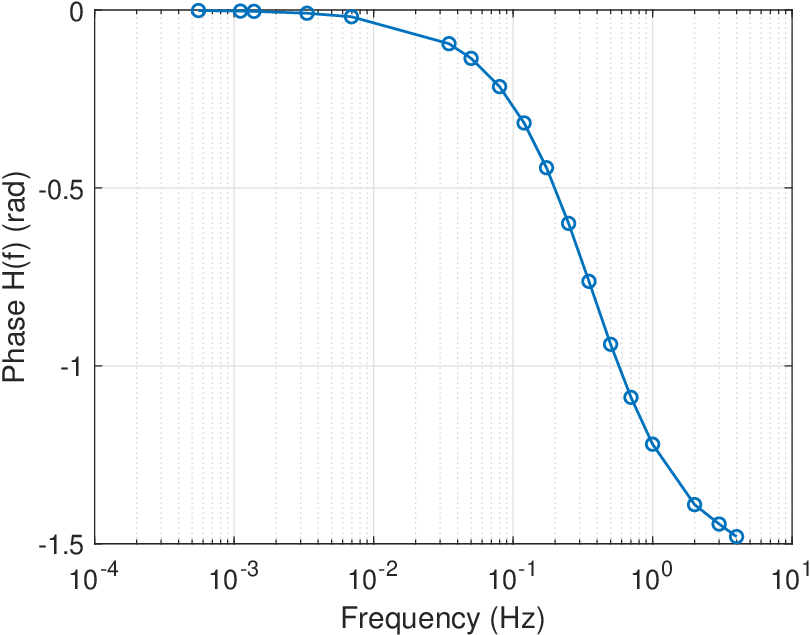}
    \label{freq_response_input_hol_to_HOL_phase}}
  \hspace{0.015\linewidth}
  \subfloat[]{
    \includegraphics[width=0.3\linewidth]{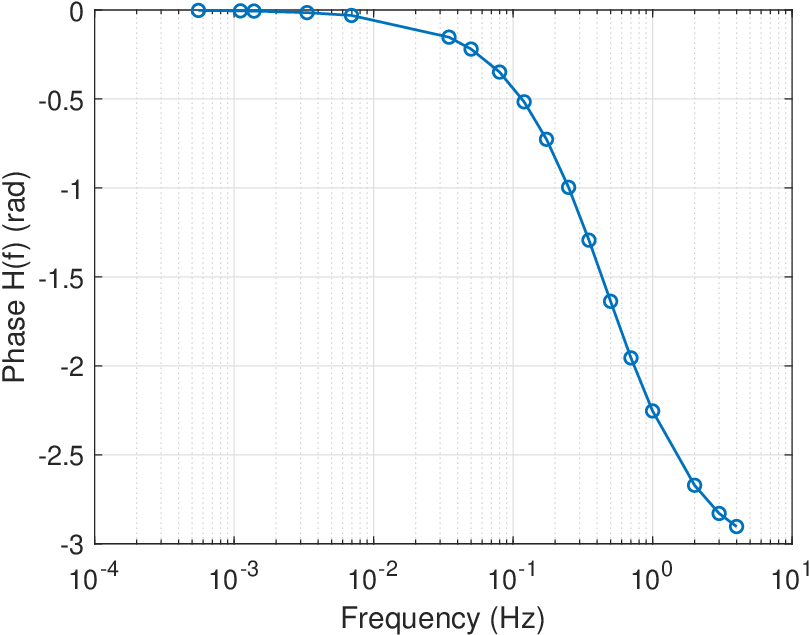}
    \label{freq_response_input_hac_to_HOL_phase}}

  \caption{Frequency-domain response of the HOL output to different odor inputs.
The top row shows the magnitude responses for HAL, HOL, and HAC inputs in
\protect\subref{freq_response_input_hal_to_HOL_magnitude},
\protect\subref{freq_response_input_hol_to_HOL_magnitude},
and \protect\subref{freq_response_input_hac_to_HOL_magnitude},
respectively. The bottom row presents the corresponding phase responses in
\protect\subref{freq_response_input_hal_to_HOL_phase},
\protect\subref{freq_response_input_hol_to_HOL_phase},
and \protect\subref{freq_response_input_hac_to_HOL_phase}.}

  \label{freq_response_HOL_all}
\end{figure*}

To study the frequency characteristics of the odor receiver, the magnitude and phase response of the system are evaluated. Since the input signals correspond to concentration levels, they cannot be negative. Therefore, the oscillating inputs are given to the system in the form shown as
\begin{equation}
\label{Oscillating_input}
\begin{split}
c_x^{air}(t) = A (1 + cos(2 \pi ft)),
\end{split}
\end{equation}

\noindent where $A = 5\times10^{-11} \, mol/m^3$ corresponds to a linear operation region according to \autoref{LinearityPilot_grid_modes_3x4}. In each simulation, only one input is made oscillatory, and the other two are taken as zero. To be able to work with oscillating outputs once there are oscillating inputs, the output of the system is considered as the internal plant HOL concentration ($c_o(t)$) in this analysis. In general, the output is of the form $c_o(t) = B (1 + cos(2 \pi ft + \phi))$, where $B$ has units $\mu mol/L$. Then, the frequency response of the system is calculated as
\begin{equation}
\label{Freq_responce}
\begin{split}
H(t) = \frac{Be^{j\phi}}{A[\mu mol/L]}.
\end{split}
\end{equation}

Using these definitions, the magnitude and phase response of the system are calculated and given in \autoref{freq_response_HOL_all}. The results show that the magnitude response remains constant at low frequencies and begins to roll off as the frequency increases for each odor molecule. This indicates that the plant's biochemical network cannot track rapid fluctuations in odor concentration. Moreover, the effective bandwidth of the biological receiver appears consistent across all three odor molecules. This suggests that the rate-limiting step in the biochemical network is likely common to all three pathways, potentially the final conversion stages leading to the HOL concentration. In addition, there is a significant difference in the magnitude response among the three inputs. HOL exhibits the highest gain, while HAC has the lowest gain. These findings confirm the observation that HOL is the dominant driver for this system.

\begin{figure}[!h]
  \centering
  \includegraphics[width=0.7\linewidth]{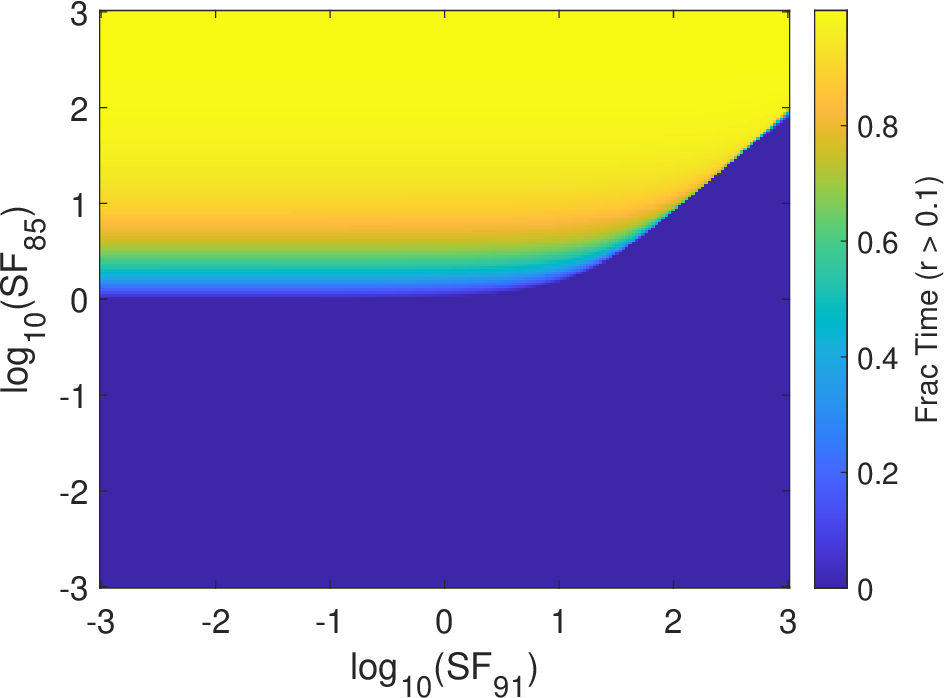}
  \caption{Sensitivity analysis of the linear operating region with respect to enzyme abundances. The horizontal and vertical axes show the scaling factors for each enzyme abundance.}
  \label{SensitHeatmap_UGT85_UGT91_mode3_T3600}
\end{figure}

To understand the sensitivity of the linear operation region to the enzyme abundances, the abundances of UGT85A53 and UGT91R1 are varied, and the linearity is examined. For this analysis, periodic on-off cycling input is applied for $1$ $hour$, and the results are given in \autoref{SensitHeatmap_UGT85_UGT91_mode3_T3600}. It can be observed that the system's linearity is significantly more sensitive to the abundance of UGT85A53. This result further demonstrates that HOL is the critical odor molecule for the receiver, as its conversion process represents the primary constraint on the system's ability to maintain a linear response.

\subsection{End-to-End Communication System}

Here, the end-to-end communication system is simulated using a single transmitter-receiver plant pair. In these simulations, equiprobable insect bites (bits) are used. The propagation channel is simulated independently for the three considered GLVs using a shared wind realization, under the assumption that the molecules do not interact or affect one another's movement. For each receiver position and for each odor molecule, (\ref{pde_diff_advec}) is solved. In addition, zero-mean white Gaussian wind is assumed for the $x$ and $y$ dimensions, while there is no wind in the $z$ dimension. In \cite{Intake_and_Transformation_to_a_Glycoside_of_Z3Hexenol_from_Infested_Neighbors_Reveals_a_Mode_of_Plant_Odor_Reception_and_Defense}, a comprehensive metabolite analysis of leaf extracts is conducted, and the amount of HEXVic was found to be $0.10 \pm 0.02 \, \mu g.g^{-1} FW$ in control plants, and $2.39 \pm 0.78 \, \mu g.g^{-1} FW$ in plants exposed to volatiles from neighboring herbivore-infested plants. In accordance with this, the alarm threshold in this article is taken as $0.5 \, \mu g.g^{-1} FW$. Using the molecular weight of HEXVic ($MW_v = 394.4 \, g/mol$), the threshold is written in units $\mu M$ as
\begin{equation}
\label{micro_g_g-1_FW_to_micro_M}
\begin{split}
c_{v0} \, [\mu M] = \frac{c_{v0} \, [\mu g.g^{-1} FW] \times 10^{-3}}{MW_v \times V_{intra}}.
\end{split}
\end{equation}

\begin{figure*}[!t]
  \centering

  \subfloat[]{
    \includegraphics[width=0.3\linewidth]{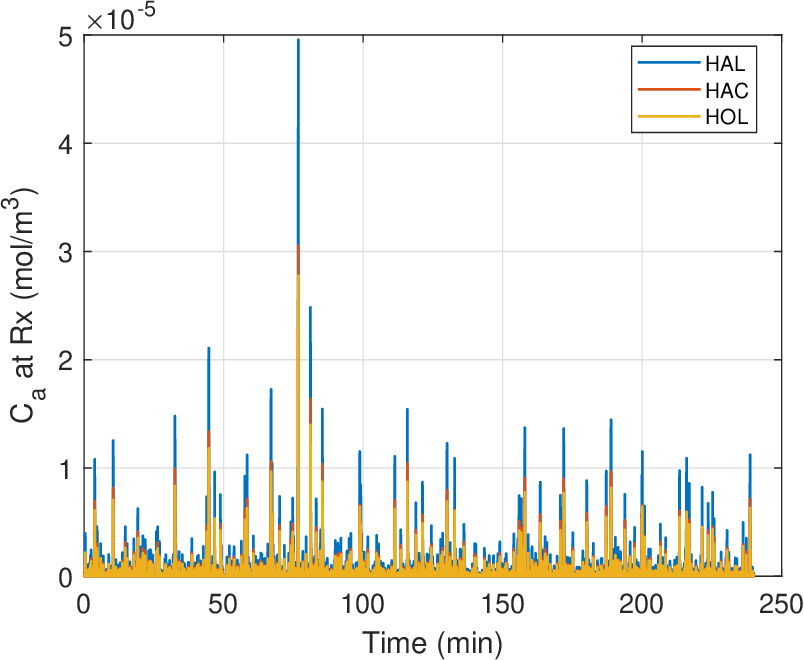}
    \label{fig_channel_rx_air_1tx1rx}}
  \hspace{0.02\linewidth}
  \subfloat[]{
    \includegraphics[width=0.3\linewidth]{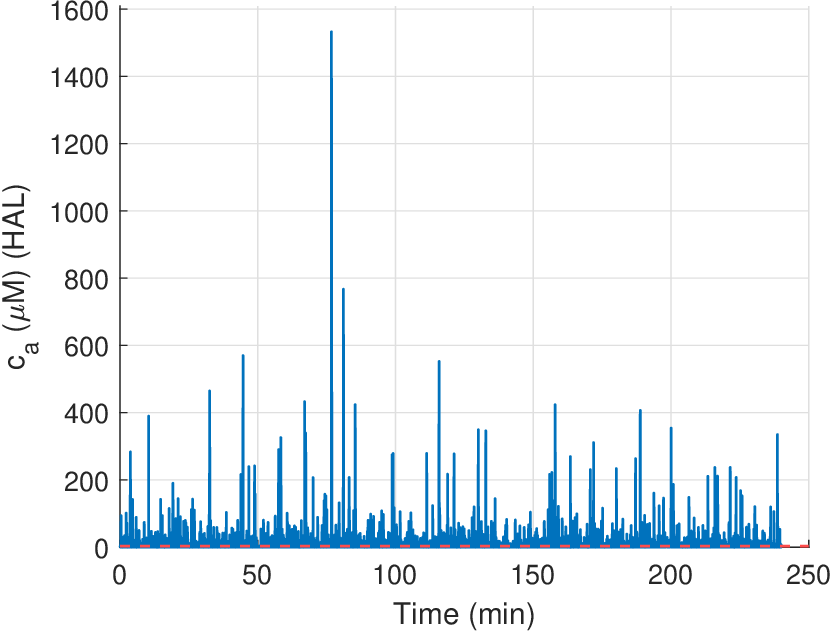}
    \label{fig_rx_c_a_hal_1tx1rx}}
  \hspace{0.02\linewidth}
  \subfloat[]{
    \includegraphics[width=0.3\linewidth]{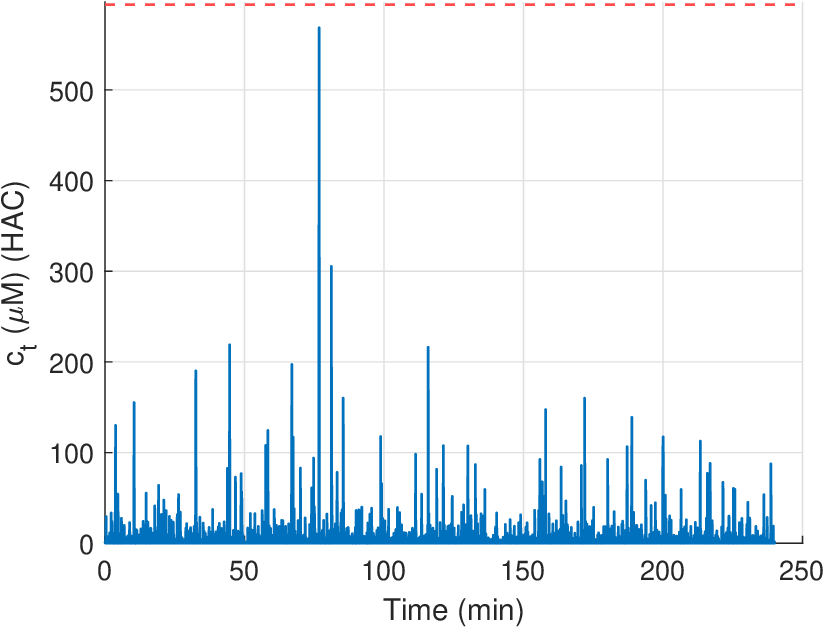}
    \label{fig_rx_c_t_hac_1tx1rx}}


  \subfloat[]{
    \includegraphics[width=0.3\linewidth]{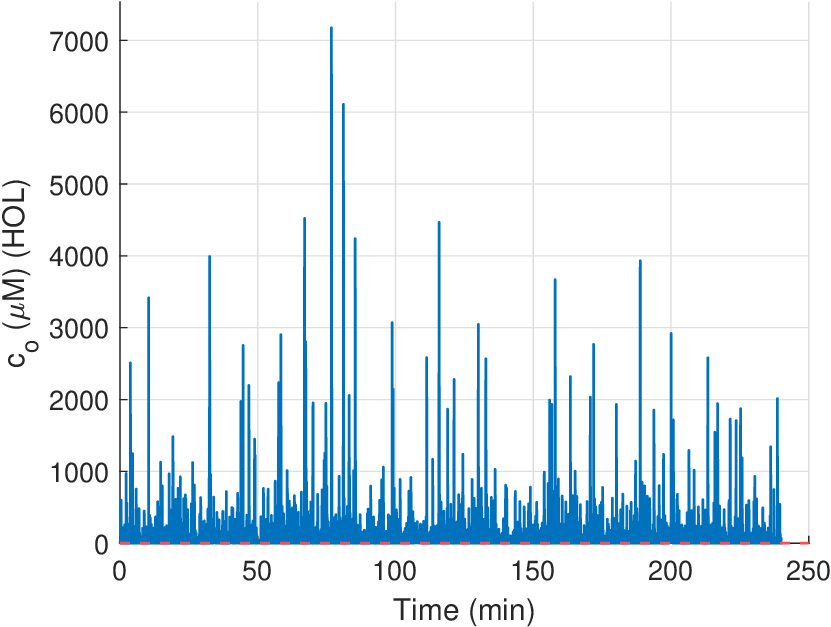}
    \label{fig_rx_c_o_hol_1tx1rx}}
  \hspace{0.02\linewidth}
  \subfloat[]{
    \includegraphics[width=0.3\linewidth]{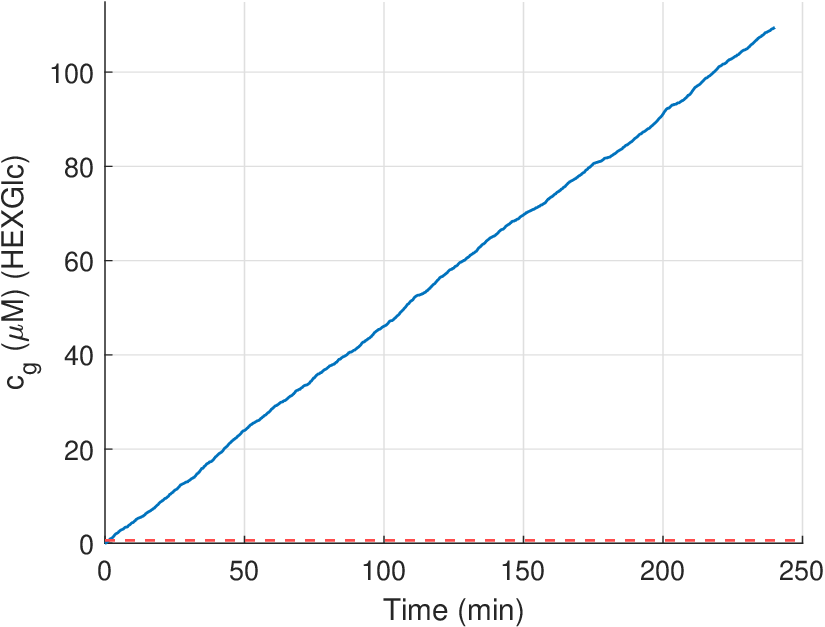}
    \label{fig_rx_c_g_hexglc_1tx1rx}}
  \hspace{0.02\linewidth}
  \subfloat[]{
    \includegraphics[width=0.3\linewidth]{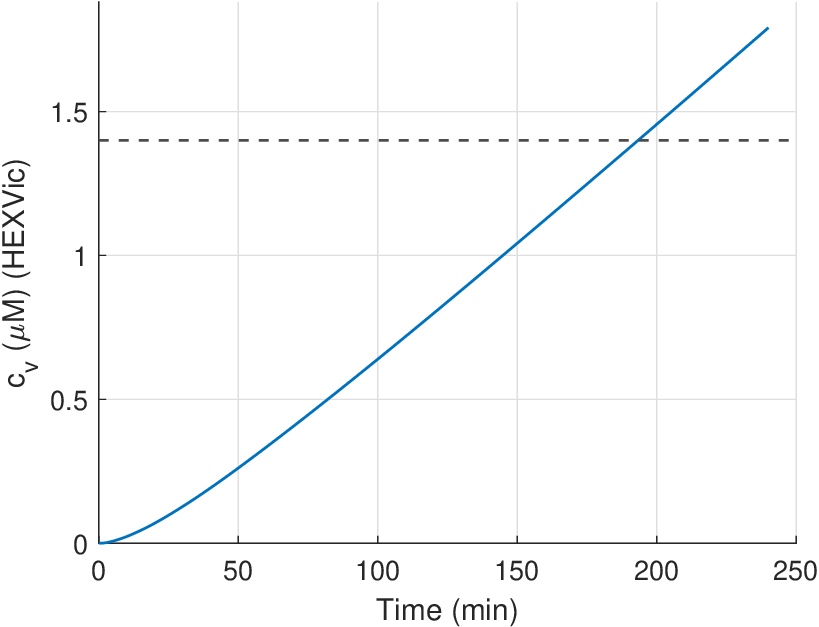}
    \label{fig_rx_c_v_hexvic_1tx1rx}}

  \caption{$1$-transmitter $1$-receiver simulation results. The channel output air concentrations are shown in \protect\subref{fig_channel_rx_air_1tx1rx}. The internal leaf concentrations are shown for \protect\subref{fig_rx_c_a_hal_1tx1rx} HAL, \protect\subref{fig_rx_c_t_hac_1tx1rx} HAC, \protect\subref{fig_rx_c_o_hol_1tx1rx} HOL, \protect\subref{fig_rx_c_g_hexglc_1tx1rx} HEXGlc, and \protect\subref{fig_rx_c_v_hexvic_1tx1rx} HEXVic. The red dashed lines indicate the linearity limit for that molecule, and the black dashed line indicates the alarm decision threshold.}
  \label{fig_system_1tx1rx_overview}
\end{figure*}

Parameters described in \autoref{simulation_params} for a non-directed strong wind scenario are used, and important signals are plotted. These results are given in \autoref{fig_system_1tx1rx_overview}. It is observed that the receiver works in the non-linear region of operation in a general natural setting. As can be seen in \autoref{fig_system_1tx1rx_overview}\protect\subref{fig_channel_rx_air_1tx1rx}, there is no bit-wise structure due to the channel effects. However, the receiver plant is able to obtain the information that "there is an attack nearby".

\begin{table}[h]
  \caption{Simulation Parameters}
  \centering
  \label{simulation_params}
  \footnotesize
  \renewcommand{\arraystretch}{1.15}
  \setlength{\tabcolsep}{5pt}
  \begin{tabular}{@{}>{\itshape}p{4.5cm}>{\upshape}p{2.7cm}@{}}
    \toprule
    \textbf{Description} & \textbf{\textit{Value}} \\
    \midrule

    \multicolumn{2}{@{}l}{\textbf{\textit{Timing and decision}}}\\
    Channel sampling rate, $f_s$ & $10$ Hz \\
    Symbol duration, $T_{\mathrm{sym}}$ & $2$ s \\
    HEXVic alarm threshold, \(\,c_{v0}\) & $1.4~\mu$M \\
    \midrule

    \multicolumn{2}{@{}l}{\textbf{\textit{Receiver-side multiplicative loss}}}\\
    Beta-loss mean, (HAL/HAC/HOL) & $0.85$ \\
    Beta-loss coefficient of variation (CV) & $0.15$ \\
    \midrule

    \multicolumn{2}{@{}l}{\textbf{\textit{Emission amplitudes}}}\\
    HAL emission amplitude, $A_a$ & $2.76\times 10^{-11}$ mol/s \\
    HOL emission amplitude, $A_o$ & $1.52\times 10^{-11}$ mol/s \\
    HAC emission amplitude, $A_t$ & $1.45\times 10^{-11}$ mol/s \\
    \midrule

    \multicolumn{2}{@{}l}{\textbf{\textit{Geometry}}}\\
    Tx position & $[0,\,0,\,1]$ m \\
    Rx position (1Tx--1Rx) & $[0.15,\,0,\,1]$ m \\
    Rx position (GLV comparison) & $[0.20,\,0,\,1]$ m \\
    \midrule

    \multicolumn{2}{@{}l}{\textbf{\textit{Wind regimes}}}\\
    Non-directed strong wind & $\mu_x = \mu_y = 0$ m/s, $\sigma_x = \sigma_y = 0.5$ m/s \\
    Non-directed weak wind &  $\mu_x = \mu_y = 0$ m/s, $\sigma_x = \sigma_y = 0.01$ m/s\\
    Directed wind &  $\mu_x = 0.2$, $\mu_y = 0$ m/s, $\sigma_x = \sigma_y = 0.01$ m/s\\
    \bottomrule
  \end{tabular}
\end{table}

\begin{figure}[!h]
  \centering
  \includegraphics[width=0.7\linewidth]{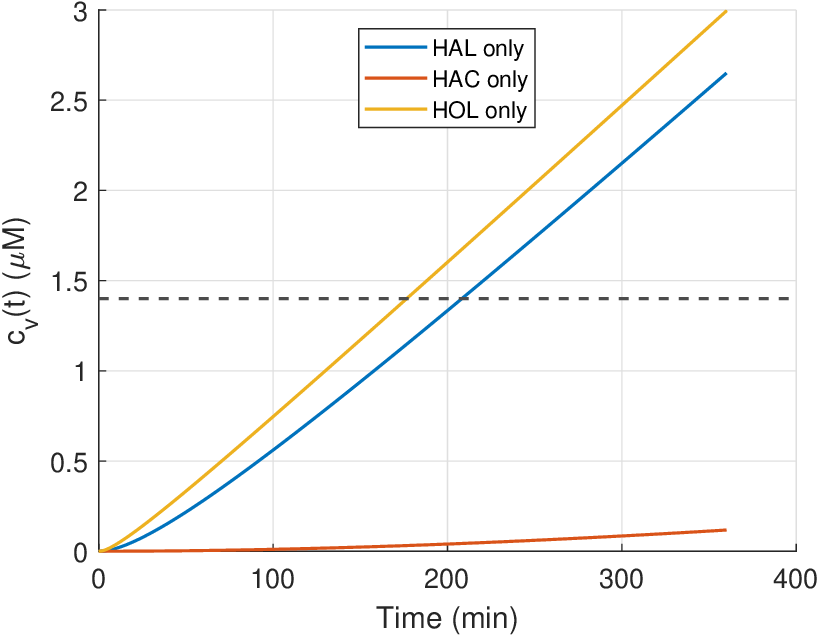}
  \caption{Comparison of internal HEXVic accumulation for single GLV emission scenarios.}
  \label{HexVic_compare_singleVOC_sameWind}
\end{figure}

From the previous receiver analysis, it was found that the input HOL air concentration is the strongest signal that contributes to the output internal HEXVic concentration. However, understanding the individual effects of HAL, HOL, and HAC from the perspective of communication engineering requires the main operations in the transmitter plant as well. In this article, the Carbon concentration budged idea is introduced. Using the emission amplitudes given in \autoref{simulation_params}, the maximum Carbon molecule emission amplitude is determined as $3.728\times 10^{-10 }\, mol/s$. Then, this amount is used to send only one type of GLV per simulation with emission amplitudes $A_a = A_o = 6.21\times 10^{-11} \, mol/s$ and $A_t = 4.66\times 10^{-11} \, mol/s$. For this analysis, the parameters in \autoref{simulation_params} for the directed wind scenario are used, and the resulting plot is given in \autoref{HexVic_compare_singleVOC_sameWind}. This result also suggests that HOL is the most efficient molecule to transmit. Then, a natural question arises: Why do plants send different GLVs, and not only HOL with higher amounts? To properly answer this question, one needs to develop a way to characterize the energy required to produce and store these molecules in a transmitter plant. Furthermore, one has to take into account the fact that storage of one molecule in large amounts can be toxic to the plant \cite{Identification_of_a_Hexenal_Reductase_That_Modulates_the_Composition_of_Green_Leaf_Volatiles}, \cite{Green_Leaf_Volatiles_A_New_Player_in_the_Protection_against_Abiotic_Stresses}. Therefore, toxicity limits for these molecules must be established. These questions motivate a multidisciplinary research direction that requires close collaboration between engineers and biologists.

One possible explanation for the usage of multiple odor molecules can be given from a communication theoretical perspective. Here, the plant is transmitting the same message over independent channels. This can be thought of as a form of diversity combining technique. We name this technique \textit{chemical diversity combining}. Chemical diversity combining does not improve the accuracy of communication by somehow overcoming the effects of the diffusion-advection channel. As can be seen in \autoref{fig_system_1tx1rx_overview}\protect\subref{fig_channel_rx_air_1tx1rx}, all odor molecules travel in the air in a similar manner. That is, the wind determines the movements. Therefore, odor molecules having different masses or diffusion coefficients do not matter in this topology. Chemical diversity combining can increase the accuracy of communication by introducing different, preferably independent, chemical channels for the same outcome. For example, insects that feed on a leaf can release salivary enzymes that suppress HOL emission from that leaf, but HAL and HAC can still be emitted. Or, the receiver plant might be in such a state that it must use HAL for different chemical reactions. In these kinds of scenarios, utilizing chemical diversity by combining techniques by sending the same message over different chemical channels can prove helpful for the communication link.

\begin{figure}[!t]
  \centering

  \subfloat[]{
    \includegraphics[width=0.7\linewidth]{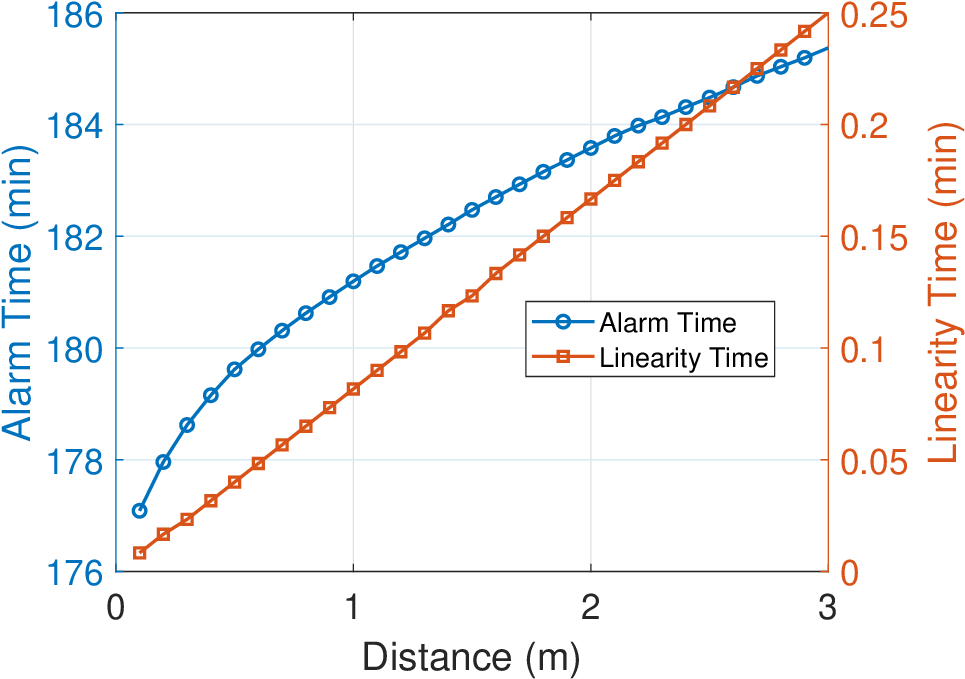}
    \label{Alarm_and_Linearity_vs_Distance_directed}}
  \hspace{0.015\linewidth}
  \subfloat[]{
    \includegraphics[width=0.7\linewidth]{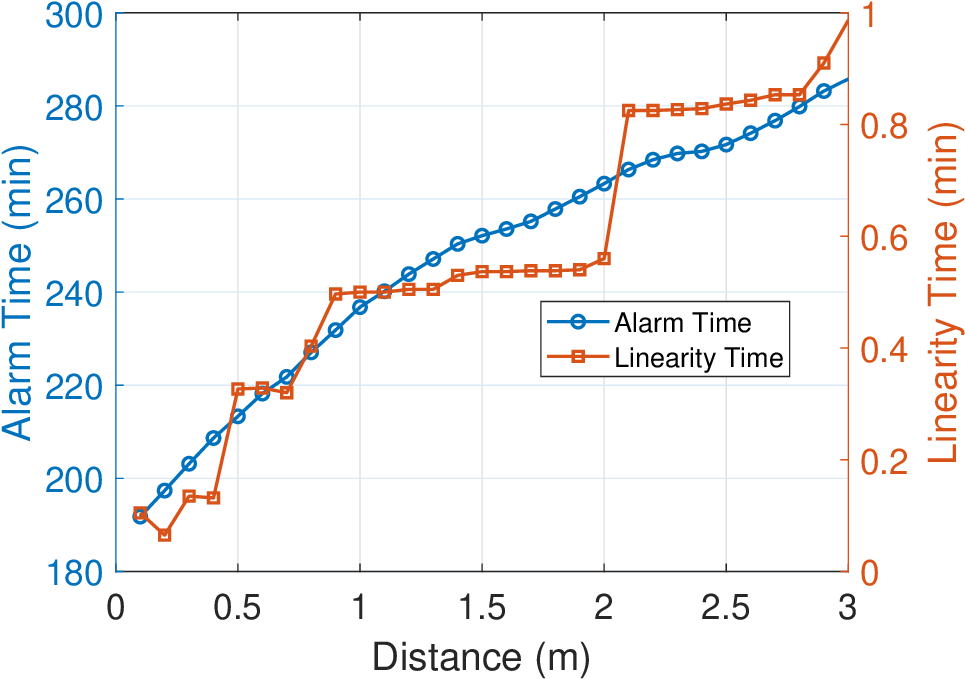}
    \label{Alarm_and_Linearity_vs_Distance_nondirected_strong}}
  \hspace{0.015\linewidth}
  \subfloat[]{
    \includegraphics[width=0.7\linewidth]{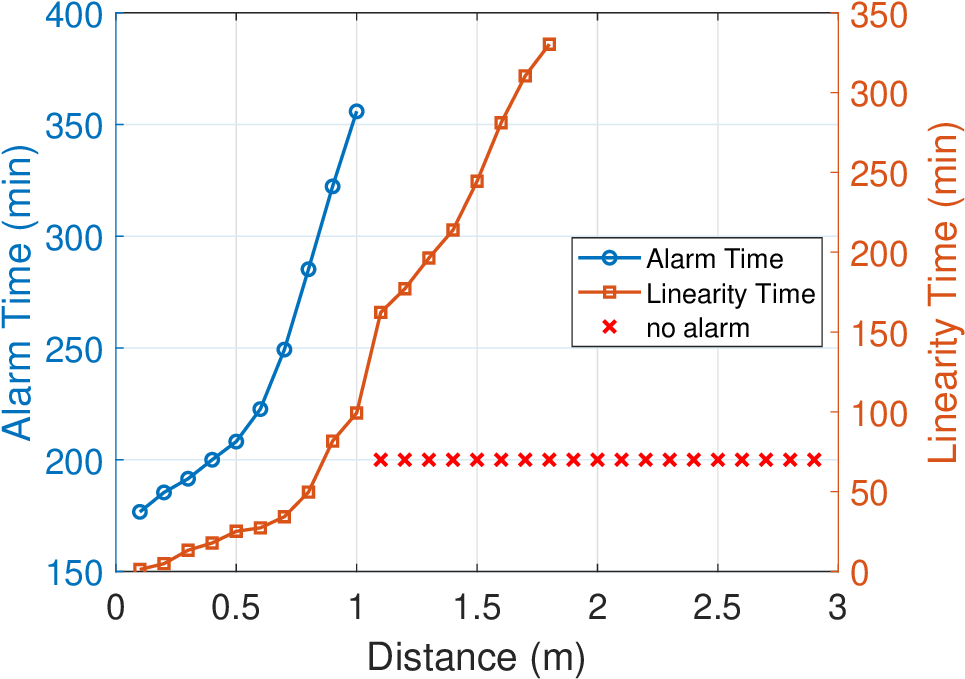}
    \label{Alarm_and_Linearity_vs_Distance_nondirected_weak}}

  \caption{Alarm time and linearity time versus Tx–Rx distance for three wind regimes:
\protect\subref{Alarm_and_Linearity_vs_Distance_directed} directed wind,
\protect\subref{Alarm_and_Linearity_vs_Distance_nondirected_strong}non-directed strong wind, and
\protect\subref{Alarm_and_Linearity_vs_Distance_nondirected_weak} non-directed weak wind.}
  \label{Time_vs_distance}
\end{figure}

Since the receiver is an accumulator, the meaningful metric to consider in this topology to quantify the communication is the time at which the receiver plant becomes "alarmed". This time is referred to as \textit{Alarm Time}. The odor communication link between a single transmitter-receiver pair is further analyzed by inspecting the alarm time in different wind scenarios at different distances. Here, distance refers to the $x$-axis position of the receiver plant. In addition, the initial time where the receiver becomes non-linear, referred to as \textit{Linearity Time}, is also inspected. The simulation results for these analyses are presented in \autoref{Time_vs_distance}. The results show that the alarm time increases with increasing distance for all wind regimes. Wind variability significantly affects this trend: while the directed-wind case yields a smooth and predictable increase, the non-directed regimes exhibit stronger distance sensitivity and non-smooth transitions. For the non-directed weak wind scenario, after $1 m$, the receiver plants do not get any message. Comparing linearity and alarm times reveals that the alarm decision typically occurs while the receiver operates in the non-linear region.

\subsection{Spread of the Alarm Message}

\begin{figure*}[!t]
  \centering

  \subfloat[]{
    \includegraphics[width=0.21\linewidth]{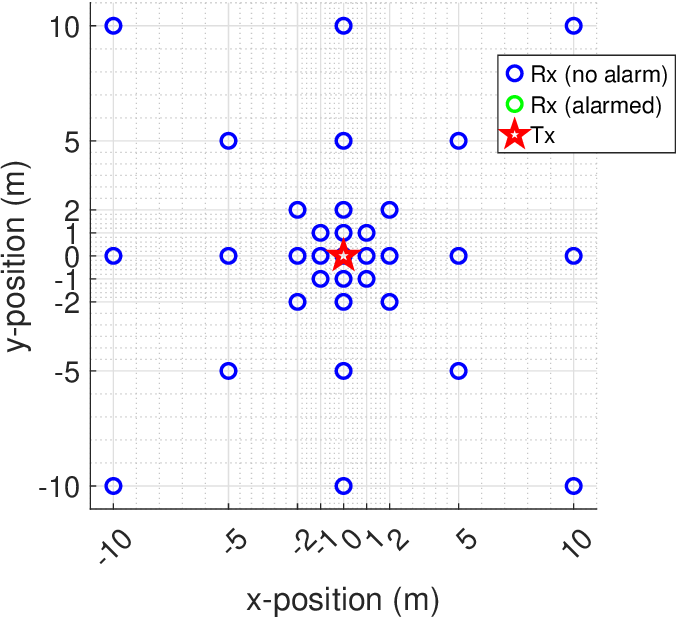}
    \label{alarmmap_strong_t01}
  } 
  \hspace{0.01\linewidth}
  \subfloat[]{
    \includegraphics[width=0.21\linewidth]{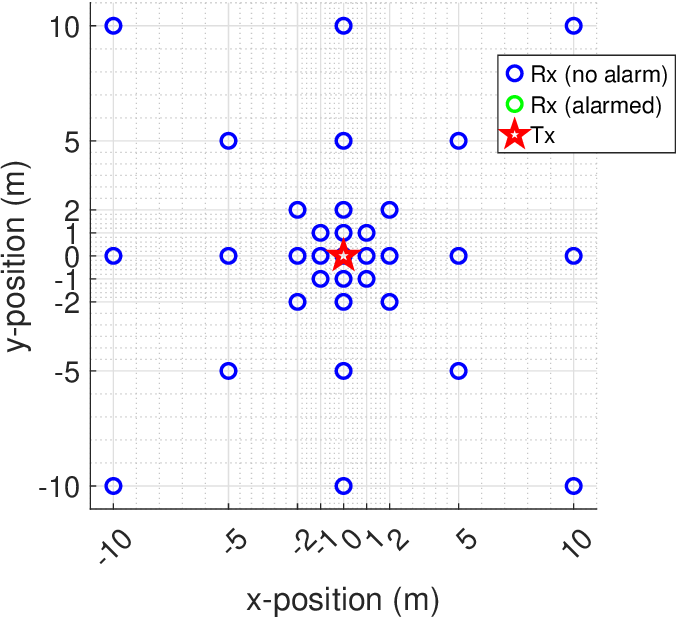}
    \label{alarmmap_strong_t03}
  }
  \hspace{0.01\linewidth}
  \subfloat[]{
    \includegraphics[width=0.21\linewidth]{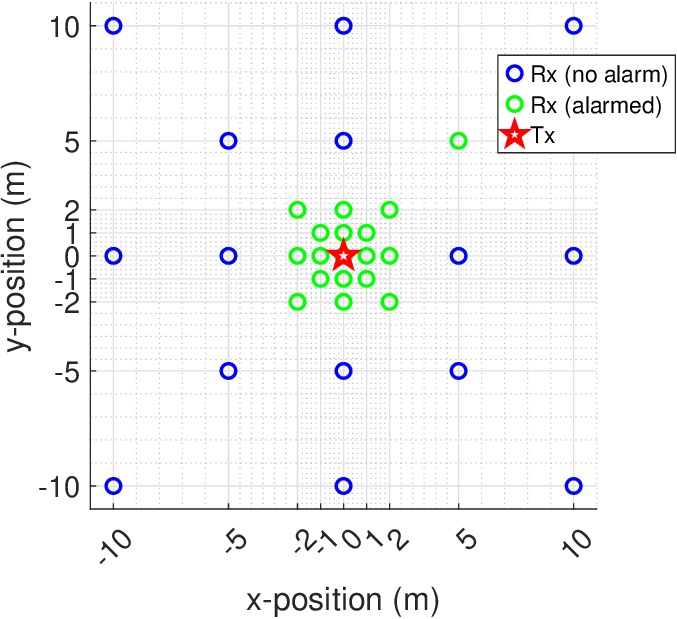}
    \label{alarmmap_strong_t06}
  }
  \hspace{0.01\linewidth}
  \subfloat[]{
    \includegraphics[width=0.21\linewidth]{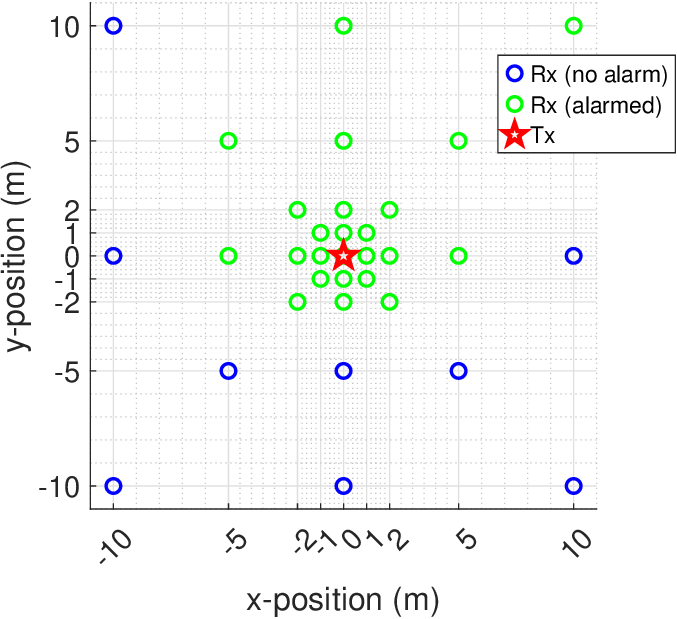}
    \label{alarmmap_strong_t10}
  }

  \vspace{0.8mm}

  \subfloat[]{
    \includegraphics[width=0.21\linewidth]{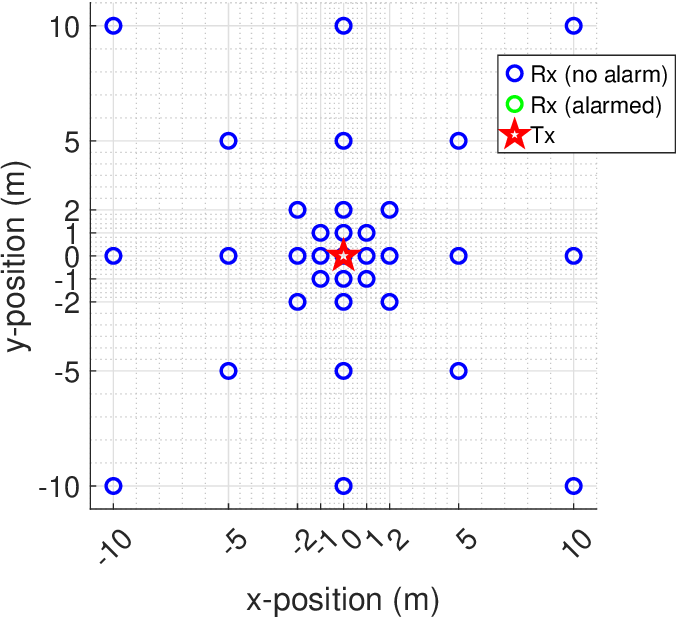}
    \label{alarmmap_weak_t01}
  }
  \hspace{0.01\linewidth}
  \subfloat[]{
    \includegraphics[width=0.21\linewidth]{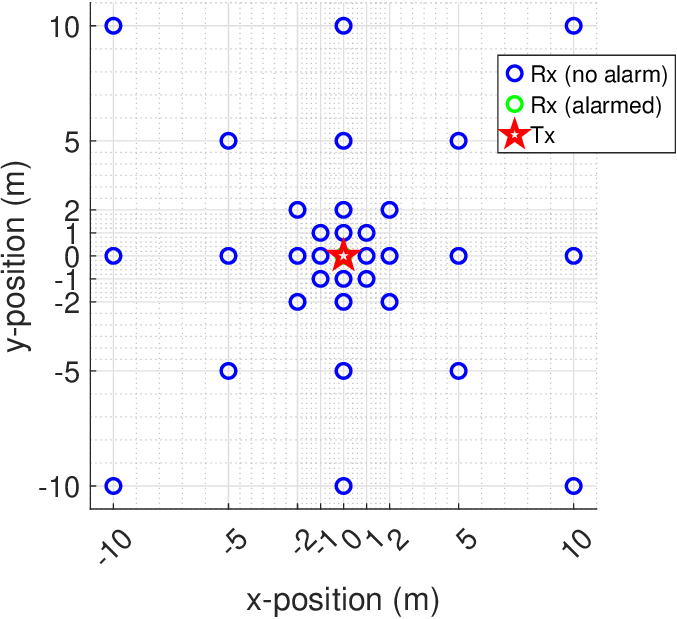}
    \label{alarmmap_weak_t03}
  }
  \hspace{0.01\linewidth}
  \subfloat[]{
    \includegraphics[width=0.21\linewidth]{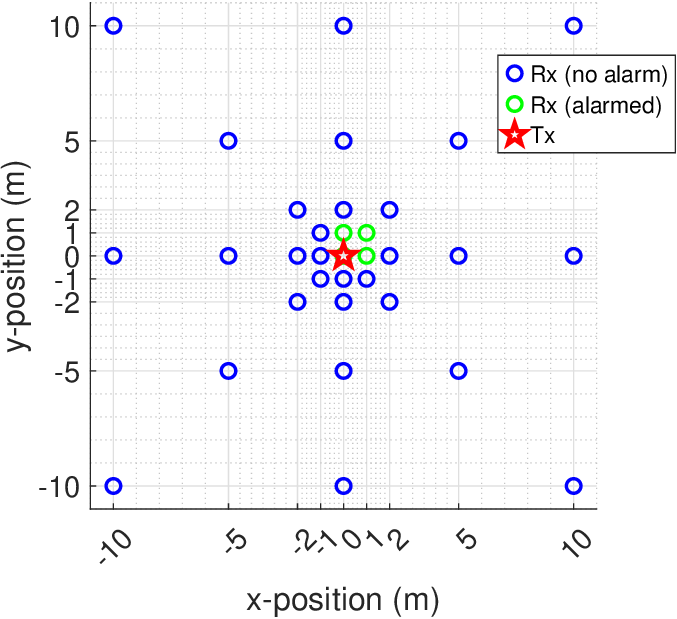}
    \label{alarmmap_weak_t06}
  }
  \hspace{0.01\linewidth}
  \subfloat[]{
    \includegraphics[width=0.21\linewidth]{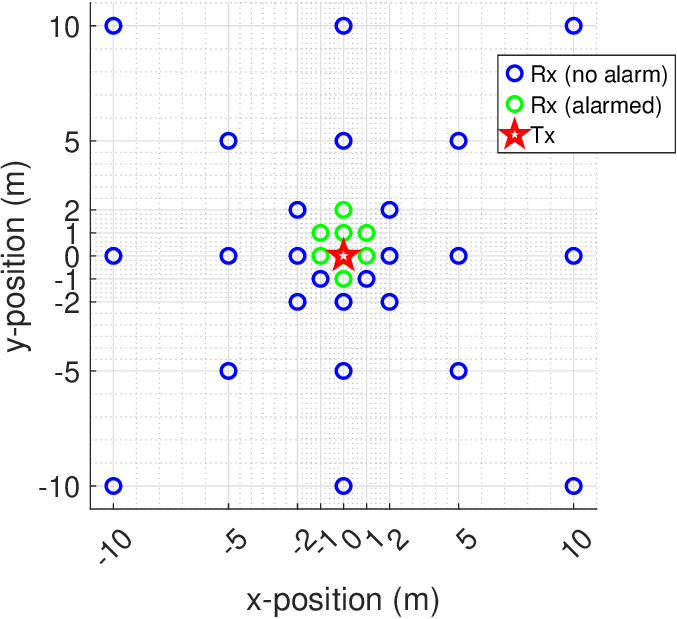}
    \label{alarmmap_weak_t10}
  }

  \caption{Population-level alarm maps under non-directed strong wind after \protect\subref{alarmmap_strong_t01} $1$ hour,  \protect\subref{alarmmap_strong_t03} $3$ hours, \protect\subref{alarmmap_strong_t06} $6$ hours, and \protect\subref{alarmmap_strong_t10} $10$ hours, and the maps under non-directed weak wind after \protect\subref{alarmmap_weak_t01} $1$ hour, \protect\subref{alarmmap_weak_t03} $3$ hours, \protect\subref{alarmmap_weak_t06} $6$ hours, and \protect\subref{alarmmap_weak_t10} $10$ hours. }
  \label{alarm_maps_weak_strong_2x4}
\end{figure*}

Here, how the warning message is received by the surrounding population is simulated. Many receiver plants are placed around one transmitter plant. The simulations are carried out for the non-directed strong wind and non-directed weak wind scenarios, assuming that the stomata of all plants are open throughout the simulation. The results are given in \autoref{alarm_maps_weak_strong_2x4}. It is observed that, under the non-directed strong wind scenario, the alarm message propagates significantly further from the transmitter. In contrast, under weak wind conditions, the alarm remains largely localized. This result, together with the results of \autoref{Time_vs_distance}, shows how the communication of the alarm message is highly dependent on the strong carrier wind. This dependency suggests a direct application in the design of automated monitoring systems for smart greenhouses. Specifically, integrating controlled airflow management can optimize signal coverage and reduce detection latency, ensuring that localized pest or pathogen outbreaks are rapidly communicated to the entire plant population and monitoring devices through the diffusion-advection channel. By maintaining a consistent medium flow, the communication radius of the biological receivers can be artificially extended, facilitating more robust and predictable bio-monitoring networks.

\section{Conclusion}

This study presents a comprehensive end-to-end communication model for odor signaling in plants, specifically focusing on the transduction of GLVs from stressed source plants to unstressed sink plants. By integrating diffusion-advection channel dynamics with a receiver-side biochemical network, the transformation of airborne (Z)-3-hexenal (HAL), (Z)-3-hexenol (HOL), and (Z)-3-hexenyl acetate (HAC) into the defensive metabolite (Z)-3-hexenyl $\beta$-vicianoside (HEXVic) is characterized. Notably, HOL was identified as the primary driver of the system and the main constraint on the receiver's linear operating region. This specific technical insight opens a deeper inquiry into the evolutionary logic of plant behavior: why has the utilization of a diverse composition of molecules evolved if a single component serves as the primary driver? By synthesizing engineering principles with biological systems, these complex signaling trade-offs can be decoded as optimized information-sharing strategies.

It is emphasized that the restriction of the stomata limits the capacity for odor communication, suggesting that future smart agriculture techniques should rely on hybrid systems that incorporate various biological signals. Although plants may not be able to decode the full bit sequence of a stress event, specialized devices could be developed to extract more detailed data, such as specific herbivore biting patterns. This hidden information could be reconstructed to allow highly informed decisions in agriculture applications.

To build on these findings, future studies should aim to uncover the mechanics of other secondary signals, such as $Ca^{2+}$ and ROS, and incorporate recovery mechanisms to model the long-term "resetting" of the biological receiver. Ultimately, the adoption of this interdisciplinary approach allows the evolutionary background of plant communication to be better understood, while a practical framework is provided to use these biological mechanisms for next-generation precision farming.

\bibliographystyle{IEEEtran}
\bibliography{references.bib}

@IEEEtranBSTCTL{IEEEexample:BSTcontrol,
  CTLuse_article_number     = "yes",
  CTLuse_paper              = "yes",
  CTLuse_url                = "yes",
  CTLuse_forced_etal        = "yes",
  CTLmax_names_forced_etal  = "6",
  CTLnames_show_etal        = "1",
  CTLuse_alt_spacing        = "yes",
  CTLalt_stretch_factor     = "4",
  CTLdash_repeated_names    = "yes",
  CTLname_format_string     = "{f.~}{vv~}{ll}{, jj}",
  CTLname_latex_cmd         = "",
  CTLname_url_prefix        = "[Online]. Available:"
}

@misc{Sustainable_and_Precision_Agriculture_with_the_Internet_of_Everything_IoE,
      title={Sustainable and Precision Agriculture with the Internet of Everything (IoE)}, 
      author={Adil Z. Babar and Ozgur B. Akan},
      year={2025},
      eprint={2404.06341},
      archivePrefix={arXiv},
      primaryClass={eess.SP},
      url={https://arxiv.org/abs/2404.06341}, 
}

@ARTICLE{Odor_Based_Molecular_Communications_State_of_the_Art_Vision_Challenges_and_Frontier_Directions,
  author={Aktas, Dilara and Ortlek, Beyza E. and Civas, Meltem and Baradari, Elham and Kilic, Ahmet B. and Bilgen, Fatih E. and Okcu, Ayse S. and Whitfield, Melanie and Cetinkaya, Oktay and Akan, Ozgur B.},
  journal={IEEE Communications Surveys \& Tutorials}, 
  title={Odor-Based Molecular Communications: State-of-the-Art, Vision, Challenges, and Frontier Directions}, 
  year={2025},
  volume={27},
  number={4},
  pages={2658-2692},
  doi={10.1109/COMST.2024.3487472}}

@article{Intake_and_Transformation_to_a_Glycoside_of_Z3Hexenol_from_Infested_Neighbors_Reveals_a_Mode_of_Plant_Odor_Reception_and_Defense,
  author = {Sugimoto, Koichi and Matsui, Kenji and Iijima, Yoko and Akakabe, Yoshihiko and Muramoto, Shoko and Ozawa, Rika and Uefune, Masayoshi and Sasaki, Ryosuke and Alamgir, Kabir Md. and Akitake, Shota and Nobuke, Tatsunori and Galis, Ivan and Aoki, Koh and Shibata, Daisuke and Takabayashi, Junji},
  doi = {10.1073/pnas.1320660111},
  journal = {Proceedings of the National Academy of Sciences},
  month = {May},
  number = {19},
  pages = {7144--7149},
  publisher = {Proceedings of the National Academy of Sciences},
  title = {Intake and Transformation to a Glycoside of ({{Z}})-3-Hexenol from Infested Neighbors Reveals a Mode of Plant Odor Reception and Defense},
  volume = {111},
  year = {2014}
}

@article{The_Carboxylesterase_AtCXE12_Converts_Volatile_Z3Hexenyl_Acetate_to_Z3Hexenol_in_Arabidopsis_Leaves,
  author = {Cofer, Tristan M and Tumlinson, James H},
  doi = {10.1093/plphys/kiaf119},
  issn = {0032-0889},
  journal = {Plant Physiology},
  month = {April},
  number = {4},
  pages = {kiaf119},
  title = {The Carboxylesterase {{AtCXE12}} Converts Volatile ({{Z}})-3-Hexenyl Acetate to ({{Z}})-3-Hexenol in {{Arabidopsis}} Leaves},
  volume = {197},
  year = {2025}
}

@article{Glucosylation_of_Z3Hexenol_Informs_Intraspecies_Interactions_in_Plants_A_Case_Study_in_Camellia_Sinensis,
  author = {Jing, Tingting and Zhang, Na and Gao, Ting and Zhao, Mingyue and Jin, Jieyang and Chen, Yongxian and Xu, Miaojing and Wan, Xiaochun and Schwab, Wilfried and Song, Chuankui},
  copyright = {{\copyright} 2018 John Wiley \& Sons Ltd},
  doi = {10.1111/pce.13479},
  issn = {1365-3040},
  journal = {Plant, Cell \& Environment},
  langid = {english},
  number = {4},
  pages = {1352--1367},
  shorttitle = {Glucosylation of ({{Z}})-3-Hexenol Informs Intraspecies Interactions in Plants},
  title = {Glucosylation of ({{Z}})-3-Hexenol Informs Intraspecies Interactions in Plants: {{A}} Case Study in {{Camellia}} Sinensis},
  volume = {42},
  year = {2019}
}

@article{Green_Leaf_Volatile_Sensory_Calcium_Transduction_in_Arabidopsis,
  author = {Aratani, Yuri and Uemura, Takuya and Hagihara, Takuma and Matsui, Kenji and Toyota, Masatsugu},
  copyright = {2023 The Author(s)},
  doi = {10.1038/s41467-023-41589-9},
  issn = {2041-1723},
  journal = {Nature Communications},
  langid = {english},
  month = {October},
  number = {1},
  pages = {6236},
  publisher = {Nature Publishing Group},
  title = {Green Leaf Volatile Sensory Calcium Transduction in {{Arabidopsis}}},
  volume = {14},
  year = {2023}
}

@article{Identification_of_a_Hexenal_Reductase_That_Modulates_the_Composition_of_Green_Leaf_Volatiles,
  author = {Tanaka, Toshiyuki and Ikeda, Ayana and Shiojiri, Kaori and Ozawa, Rika and Shiki, Kazumi and {Nagai-Kunihiro}, Naoko and Fujita, Kenya and Sugimoto, Koichi and Yamato, Katsuyuki T. and Dohra, Hideo and Ohnishi, Toshiyuki and Koeduka, Takao and Matsui, Kenji},
  doi = {10.1104/pp.18.00632},
  issn = {0032-0889},
  journal = {Plant Physiology},
  month = {October},
  number = {2},
  pages = {552--564},
  title = {Identification of a {{Hexenal Reductase That Modulates}} the {{Composition}} of {{Green Leaf Volatiles}}},
  volume = {178},
  year = {2018}
}

@article{Identification_of_a_Tomato_UDParabinosyltransferase_for_Airborne_Volatile_Reception,
  author = {Sugimoto, Koichi and Ono, Eiichiro and Inaba, Tamaki and Tsukahara, Takehiko and Matsui, Kenji and Horikawa, Manabu and Toyonaga, Hiromi and Fujikawa, Kohki and Osawa, Tsukiho and Homma, Shunichi and Kiriiwa, Yoshikazu and Ohmura, Ippei and Miyagawa, Atsushi and Yamamura, Hatsuo and Fujii, Mikio and Ozawa, Rika and Watanabe, Bunta and Miura, Kenji and Ezura, Hiroshi and Ohnishi, Toshiyuki and Takabayashi, Junji},
  copyright = {2023 The Author(s)},
  doi = {10.1038/s41467-023-36381-8},
  issn = {2041-1723},
  journal = {Nature Communications},
  langid = {english},
  month = {February},
  number = {1},
  pages = {677},
  publisher = {Nature Publishing Group},
  title = {Identification of a Tomato {{UDP-arabinosyltransferase}} for Airborne Volatile Reception},
  volume = {14},
  year = {2023}
}

@article{Metabolomics_Reveal_Induction_of_ROS_Production_and_Glycosylation_Events_in_Wheat_Upon_Exposure_to_the_Green_Leaf_Volatile_Z3Hexenyl_Acetate,
  author = {Ameye, Maarten and Van Meulebroek, Lieven and Meuninck, Bianca and Vanhaecke, Lynn and Smagghe, Guy and Haesaert, Geert and Audenaert, Kris},
  doi = {10.3389/fpls.2020.596271},
  issn = {1664-462X},
  journal = {Frontiers in Plant Science},
  langid = {english},
  month = {December},
  publisher = {Frontiers},
  title = {Metabolomics {{Reveal Induction}} of {{ROS Production}} and {{Glycosylation Events}} in {{Wheat Upon Exposure}} to the {{Green Leaf Volatile Z-3-Hexenyl Acetate}}},
  volume = {11},
  year = {2020}
}

@article{An_Absorption_Model_of_Volatile_Organic_Compound_by_Plant_Leaf_The_Most_Influential_Site_in_the_Absorption_Pathway,
  author = {Yamane, Mizuki and Tani, Akira},
  doi = {10.1016/j.aeaoa.2024.100274},
  issn = {2590-1621},
  journal = {Atmospheric Environment: X},
  month = {August},
  pages = {100274},
  shorttitle = {An Absorption Model of Volatile Organic Compound by Plant Leaf},
  title = {An Absorption Model of Volatile Organic Compound by Plant Leaf: {{The}} Most Influential Site in the Absorption Pathway},
  volume = {23},
  year = {2024}
}

@incollection{Uptake_and_Conversion_of_Volatile_Compounds_in_PlantPlant_Communication,
  address = {Cham},
  author = {Sugimoto, Koichi and Matsui, Kenji and Takabayashi, Junji},
  booktitle = {Deciphering {{Chemical Language}} of {{Plant Communication}}},
  doi = {10.1007/978-3-319-33498-1_13},
  editor = {Blande, James D. and Glinwood, Robert},
  isbn = {978-3-319-33498-1},
  langid = {english},
  pages = {305--316},
  publisher = {Springer International Publishing},
  title = {Uptake and {{Conversion}} of {{Volatile Compounds}} in {{Plant}}--{{Plant Communication}}},
  year = {2016}
}

@article{Water_Deficit_Enhances_C_Export_to_the_Roots_in_Arabidopsis_Thaliana_Plants_with_Contribution_of_Sucrose_Transporters_in_Both_Shoot_and_Roots1OPEN,
  author = {Durand, Micka{\"e}l and Porcheron, Beno{\^i}t and Hennion, Nils and Maurousset, Laurence and Lemoine, R{\'e}mi and Pourtau, Nathalie},
  doi = {10.1104/pp.15.01926},
  issn = {0032-0889},
  journal = {Plant Physiology},
  month = {March},
  number = {3},
  pages = {1460--1479},
  pmcid = {PMC4775148},
  pmid = {26802041},
  title = {Water {{Deficit Enhances C Export}} to the {{Roots}} in {{Arabidopsis}} Thaliana {{Plants}} with {{Contribution}} of {{Sucrose Transporters}} in {{Both Shoot}} and {{Roots1}}[{{OPEN}}]},
  volume = {170},
  year = {2016}
}

@article{The_Arabidopsis_Leaf_Quantitative_Atlas_A_Cellular_and_Subcellular_Mapping_through_Unified_Data_Integration,
  author = {Tolleter, Dimitri and Smith, Edward N. and {Dupont-Thibert}, Cl{\'e}mence and Uwizeye, Clarisse and Vile, Denis and Gloaguen, Pauline and Falconet, Denis and Finazzi, Giovanni and Vandenbrouck, Yves and Curien, Gilles},
  doi = {10.1017/qpb.2024.1},
  issn = {2632-8828},
  journal = {Quantitative Plant Biology},
  month = {February},
  pages = {e2},
  pmcid = {PMC10988163},
  pmid = {38572078},
  shorttitle = {The {{Arabidopsis}} Leaf Quantitative Atlas},
  title = {The {{Arabidopsis}} Leaf Quantitative Atlas: A Cellular and Subcellular Mapping through Unified Data Integration},
  volume = {5},
  year = {2024}
}

@article{Green_Leaf_Volatiles_in_the_AtmosphereProperties_Transformation_and_Significance,
  article-number = {1655},
  author = {Sarang, Kumar and Rudziński, Krzysztof J. and Szmigielski, Rafał},
  doi = {10.3390/atmos12121655},
  issn = {2073-4433},
  journal = {Atmosphere},
  number = {12},
  title = {Green Leaf Volatiles in the Atmosphere—Properties, Transformation, and Significance},
  url = {https://www.mdpi.com/2073-4433/12/12/1655},
  volume = {12},
  year = {2021}
}

@book{The_Properties_of_Gases_and_Liquids,
  address = {New York},
  author = {Poling, Bruce E. and Prausnitz, J. M. and O'Connell, John P.},
  edition = {5th ed},
  isbn = {978-0-07-149999-6},
  langid = {english},
  lccn = {TP242},
  publisher = {McGraw-Hill},
  title = {The Properties of Gases and Liquids},
  year = {2001}
}

@book{Atmospheric_Chemistry_and_Physics_From_Air_Pollution_to_Climate_Change,
  address = {New York},
  author = {Seinfeld, John H. and Pandis, Spyros N.},
  edition = {2nd ed},
  isbn = {978-0-471-72017-1},
  langid = {english},
  lccn = {551.511},
  publisher = {J. Wiley \& sons},
  shorttitle = {Atmospheric Chemistry and Physics},
  title = {Atmospheric Chemistry and Physics: From Air Pollution to Climate Change},
  year = {2006}
}

@article{Protein_Abundance_Biases_the_Amino_Acid_Composition_of_Disordered_Regions_to_Minimize_Nonfunctional_Interactions,
  author = {Dubreuil, Benjamin and Matalon, Or and Levy, Emmanuel D.},
  doi = {10.1016/j.jmb.2019.08.008},
  issn = {0022-2836},
  journal = {Journal of Molecular Biology},
  month = {December},
  number = {24},
  pages = {4978--4992},
  title = {Protein {{Abundance Biases}} the {{Amino Acid Composition}} of {{Disordered Regions}} to {{Minimize Non-functional Interactions}}},
  volume = {431},
  year = {2019}
}

@article{Channel_Modeling_for_Diffusive_Molecular_CommunicationA_Tutorial_Review,
  author = {Jamali, Vahid and Ahmadzadeh, Arman and Wicke, Wayan and Noel, Adam and Schober, Robert},
  doi = {10.1109/JPROC.2019.2919455},
  issn = {1558-2256},
  journal = {Proceedings of the IEEE},
  month = {July},
  number = {7},
  pages = {1256--1301},
  title = {Channel {{Modeling}} for {{Diffusive Molecular Communication}}---{{A Tutorial Review}}},
  volume = {107},
  year = {2019}
}

@misc{Information_and_Communication_Theoretical_Foundations_of_the_Internet_of_Plants_Principles_Challenges_and_Future_Directions,
      title={Information and Communication Theoretical Foundations of the Internet of Plants, Principles, Challenges, and Future Directions}, 
      author={Ahmet B. Kilic and Ozgur B. Akan},
      year={2025},
      eprint={2509.08434},
      archivePrefix={arXiv},
      primaryClass={eess.SP},
      url={https://arxiv.org/abs/2509.08434}, 
}

@ARTICLE{EndtoEnd_Mathematical_Modeling_of_Stress_Communication_Between_Plants,
  author={Kilic, Ahmet Burak and Akan, Ozgur B.},
  journal={IEEE Transactions on Molecular, Biological, and Multi-Scale Communications}, 
  title={End-to-End Mathematical Modeling of Stress Communication Between Plants}, 
  year={2026},
  volume={12},
  number={},
  pages={69-78},
  doi={10.1109/TMBMC.2025.3626218}}

@article{Green_Leaf_Volatiles_A_New_Player_in_the_Protection_against_Abiotic_Stresses,
  author = {Engelberth, Jurgen},
  copyright = {http://creativecommons.org/licenses/by/3.0/},
  doi = {10.3390/ijms25179471},
  issn = {1422-0067},
  journal = {International Journal of Molecular Sciences},
  langid = {english},
  month = {January},
  number = {17},
  pages = {9471},
  publisher = {Multidisciplinary Digital Publishing Institute},
  shorttitle = {Green {{Leaf Volatiles}}},
  title = {Green {{Leaf Volatiles}}: {{A New Player}} in the {{Protection}} against {{Abiotic Stresses}}?},
  volume = {25},
  year = {2024}
}

@article{Green_Leaf_VolatilesThe_Forefront_of_Plant_Responses_Against_Biotic_Attack,
  author = {Matsui, Kenji and Engelberth, Jurgen},
  doi = {10.1093/pcp/pcac117},
  issn = {1471-9053},
  journal = {Plant and Cell Physiology},
  month = {October},
  number = {10},
  pages = {1378--1390},
  title = {Green {{Leaf Volatiles}}---{{The Forefront}} of {{Plant Responses Against Biotic Attack}}},
  volume = {63},
  year = {2022}
}

@article{Volatile_CompoundMediated_PlantPlant_Interactions_under_Stress_with_the_Tea_Plant_as_a_Model,
  author = {Jin, Jieyang and Zhao, Mingyue and Jing, Tingting and Zhang, Mengting and Lu, Mengqian and Yu, Guomeng and Wang, Jingming and Guo, Danyang and Pan, Yuting and Hoffmann, Timothy D and Schwab, Wilfried and Song, Chuankui},
  doi = {10.1093/hr/uhad143},
  issn = {2662-6810},
  journal = {Horticulture Research},
  month = {September},
  number = {9},
  pages = {uhad143},
  title = {Volatile Compound-Mediated Plant--Plant Interactions under Stress with the Tea Plant as a Model},
  volume = {10},
  year = {2023}
}

@article{Complex_Odor_from_Plants_under_Attack_Herbivores_Enemies_React_to_the_Whole_Not_Its_Parts,
  author = {van Wijk, Michiel and de Bruijn, Paulien J. A. and Sabelis, Maurice W.},
  doi = {10.1371/journal.pone.0021742},
  issn = {1932-6203},
  journal = {PLOS ONE},
  langid = {english},
  month = {July},
  number = {7},
  pages = {e21742},
  publisher = {Public Library of Science},
  shorttitle = {Complex {{Odor}} from {{Plants}} under {{Attack}}},
  title = {Complex {{Odor}} from {{Plants}} under {{Attack}}: {{Herbivore}}'s {{Enemies React}} to the {{Whole}}, {{Not Its Parts}}},
  volume = {6},
  year = {2011}
}

@article{StressInduced_Volatile_Emissions_and_Signalling_in_InterPlant_Communication,
  author = {Midzi, Joanah and Jeffery, David W. and Baumann, Ute and Rogiers, Suzy and Tyerman, Stephen D. and Pagay, Vinay},
  copyright = {http://creativecommons.org/licenses/by/3.0/},
  doi = {10.3390/plants11192566},
  issn = {2223-7747},
  journal = {Plants},
  langid = {english},
  month = {January},
  number = {19},
  pages = {2566},
  publisher = {Multidisciplinary Digital Publishing Institute},
  title = {Stress-{{Induced Volatile Emissions}} and {{Signalling}} in {{Inter-Plant Communication}}},
  volume = {11},
  year = {2022}
}

@article{Plant_Communication_across_Different_Environmental_Contexts_Suggests_a_Role_for_Stomata_in_Volatile_Perception,
  author = {Aguirre, Natalie M. and Grunseich, John M. and Lima, Andre{\'i}sa F. and Davis, Stephen D. and Helms, Anjel M.},
  copyright = {\copyright{} 2023 The Authors. Plant, Cell \& Environment published by John Wiley \& Sons Ltd.},
  doi = {10.1111/pce.14601},
  issn = {1365-3040},
  journal = {Plant, Cell \& Environment},
  langid = {english},
  number = {7},
  pages = {2017--2030},
  title = {Plant Communication across Different Environmental Contexts Suggests a Role for Stomata in Volatile Perception},
  volume = {46},
  year = {2023}
}

@online{chemeo_hexenal,
  author       = {{Chemeo Database}},
  title        = {cis-3-Hexenal},
  url          = {https://www.chemeo.com/cid/46-326-9/3-Hexenal-Z},
  note         = {Accessed: Sept. 21, 2025}
}

@online{chemeo_hexenol,
  author       = {{Chemeo Database}},
  title        = {cis-3-Hexen-1-ol},
  url          = {https://www.chemeo.com/cid/18-655-5/3-Hexen-1-ol-Z},
  note         = {Accessed: Sept. 21, 2025}
}

@online{chemeo_hexenylacetate,
  author       = {{Chemeo Database}},
  title        = {cis-3-Hexenyl Acetate},
  url          = {https://www.chemeo.com/cid/54-175-8/3-Hexen-1-ol-acetate-Z},
  note         = {Accessed: Sept. 21, 2025}
}

@online{epa_conversion,
  author       = {{U.S. Environmental Protection Agency}},
  title        = {Indoor Air Unit Conversions},
  url          = {https://www3.epa.gov/ceampubl/learn2model/part-two/onsite/doc/Indoor%20Air%20Unit%20Conversions.pdf},
  note         = {Accessed: Sept. 21, 2025}
}

@book{Enzyme_Kinetics_Behavior_and_Analysis_of_Rapid_Equilibrium_and_SteadyState_Enzyme_Systems_National_Library_of_Medicine_Institution,
  title={Enzyme Kinetics: Behavior and Analysis of Rapid Equilibrium and Steady-State Enzyme Systems},
  author={Segel, Irwin H.},
  year={1975},
  publisher={Wiley},
  address={New York}
}

@online{paxdb,
  author       = {{PaxDb Consortium}},
  title        = {PaxDb: Protein Abundance Database},
  url          = {https://pax-db.org},
  note         = {Accessed: Sept. 24, 2025}
}

@misc{BioRender_Merdan_2026_3,
  author       = {Merdan, Fatih},
  title        = {Created in BioRender},
  year         = {2026},
  howpublished = {\url{https://BioRender.com/z55x3op}},
  note         = {BioRender Illustration}
}

@misc{Airborne_Particle_Communication_Through_Timevarying_DiffusionAdvection_Channels,
      title={Airborne Particle Communication Through Time-varying Diffusion-Advection Channels}, 
      author={Fatih Merdan and Ozgur B. Akan},
      year={2026},
      eprint={2601.08534},
      archivePrefix={arXiv},
      primaryClass={eess.SP},
      url={https://arxiv.org/abs/2601.08534}, 
}

@article{Silencing_the_Alarm_An_Insect_Salivary_Enzyme_Closes_Plant_Stomata_and_Inhibits_Volatile_Release,
  author = {Lin, Po-An and Chen, Yintong and {Chaverra-Rodriguez}, Duverney and Heu, Chan Chin and Zainuddin, Nursyafiqi Bin and Sidhu, Jagdeep Singh and Peiffer, Michelle and Tan, Ching-Wen and Helms, Anjel and Kim, Donghun and Ali, Jared and Rasgon, Jason L. and Lynch, Jonathan and Anderson, Charles T. and Felton, Gary W.},
  copyright = {\copyright{} 2021 The Authors. New Phytologist \copyright{} 2021 New Phytologist Foundation},
  doi = {10.1111/nph.17214},
  issn = {1469-8137},
  journal = {New Phytologist},
  langid = {english},
  number = {2},
  pages = {793--803},
  shorttitle = {Silencing the Alarm},
  title = {Silencing the Alarm: An Insect Salivary Enzyme Closes Plant Stomata and Inhibits Volatile Release},
  volume = {230},
  year = {2021}
}

@article{Performance_of_Some_Estimators_of_Relative_Variability,
  author = {Ospina, Raydonal and {Marmolejo-Ramos}, Fernando},
  doi = {10.3389/fams.2019.00043},
  issn = {2297-4687},
  journal = {Frontiers in Applied Mathematics and Statistics},
  langid = {english},
  month = {August},
  publisher = {Frontiers},
  title = {Performance of {{Some Estimators}} of {{Relative Variability}}},
  volume = {5},
  year = {2019}
}

@article{Simulation_and_Inference_Algorithms_for_Stochastic_Biochemical_Reaction_Networks_From_Basic_Concepts_to_StateoftheArt,
  author = {Warne, David J. and Baker, Ruth E. and Simpson, Matthew J.},
  doi = {10.1098/rsif.2018.0943},
  issn = {1742-5689},
  journal = {Journal of The Royal Society Interface},
  month = {February},
  number = {151},
  pages = {20180943},
  shorttitle = {Simulation and Inference Algorithms for Stochastic Biochemical Reaction Networks},
  title = {Simulation and Inference Algorithms for Stochastic Biochemical Reaction Networks: From Basic Concepts to State-of-the-Art},
  volume = {16},
  year = {2019}
}

@article{Defence_Priming_in_Arabidopsis_a_MetaAnalysis,
  author = {Westman, Sara M. and Kloth, Karen J. and Hanson, Johannes and Ohlsson, Anna B. and Albrectsen, Benedicte R.},
  copyright = {2019 The Author(s)},
  doi = {10.1038/s41598-019-49811-9},
  issn = {2045-2322},
  journal = {Scientific Reports},
  langid = {english},
  month = {September},
  number = {1},
  pages = {13309},
  publisher = {Nature Publishing Group},
  title = {Defence Priming in {{Arabidopsis}} -- a {{Meta-Analysis}}},
  volume = {9},
  year = {2019}
}

@article{Herbivorous_Caterpillars_and_the_Green_Leaf_Volatile_GLV_Quandary,
  author = {Jones, Anne C. and Cofer, Tristan M. and Engelberth, Jurgen and Tumlinson, James H.},
  doi = {10.1007/s10886-021-01330-6},
  issn = {1573-1561},
  journal = {Journal of Chemical Ecology},
  langid = {english},
  month = {March},
  number = {3},
  pages = {337--345},
  title = {Herbivorous {{Caterpillars}} and the {{Green Leaf Volatile}} ({{GLV}}) {{Quandary}}},
  volume = {48},
  year = {2022}
}

@article{How_Do_Plants_Sense_Volatiles_Sent_by_Other_Plants,
  author = {Loreto, Francesco and D'Auria, Sabato},
  doi = {10.1016/j.tplants.2021.08.009},
  issn = {1360-1385},
  journal = {Trends in Plant Science},
  month = {January},
  number = {1},
  pages = {29--38},
  title = {How Do Plants Sense Volatiles Sent by Other Plants?},
  volume = {27},
  year = {2022}
}

@article{PlantPlant_Communication_Is_There_a_Role_for_Volatile_DamageAssociated_Molecular_Patterns,
  author = {Meents, Anja K. and Mith{\"o}fer, Axel},
  doi = {10.3389/fpls.2020.583275},
  issn = {1664-462X},
  journal = {Frontiers in Plant Science},
  langid = {english},
  month = {October},
  publisher = {Frontiers},
  shorttitle = {Plant--{{Plant Communication}}},
  title = {Plant--{{Plant Communication}}: {{Is There}} a {{Role}} for {{Volatile Damage-Associated Molecular Patterns}}?},
  volume = {11},
  year = {2020}
}

@article{An_EndtoEnd_Model_of_Plant_Pheromone_Channel_for_Long_Range_Molecular_Communication,
  author = {Unluturk, Bige D. and Akyildiz, Ian F.},
  doi = {10.1109/TNB.2016.2628047},
  issn = {1558-2639},
  journal = {IEEE Transactions on NanoBioscience},
  month = {January},
  number = {1},
  pages = {11--20},
  title = {An {{End-to-End Model}} of {{Plant Pheromone Channel}} for {{Long Range Molecular Communication}}},
  volume = {16},
  year = {2017}
}

@misc{Modeling_and_Analysis_of_VOCbased_Interplant_Molecular_Communication_Channel,
      title={Modeling and Analysis of VOC-based Interplant Molecular Communication Channel}, 
      author={Bitop Maitra and Ozgur B. Akan},
      year={2025},
      eprint={2512.12035},
      archivePrefix={arXiv},
      primaryClass={eess.SP},
      url={https://arxiv.org/abs/2512.12035}, 
}

\vspace{11pt}

\begin{IEEEbiography}[{\includegraphics[width=1in,height=1.25in,clip,keepaspectratio]{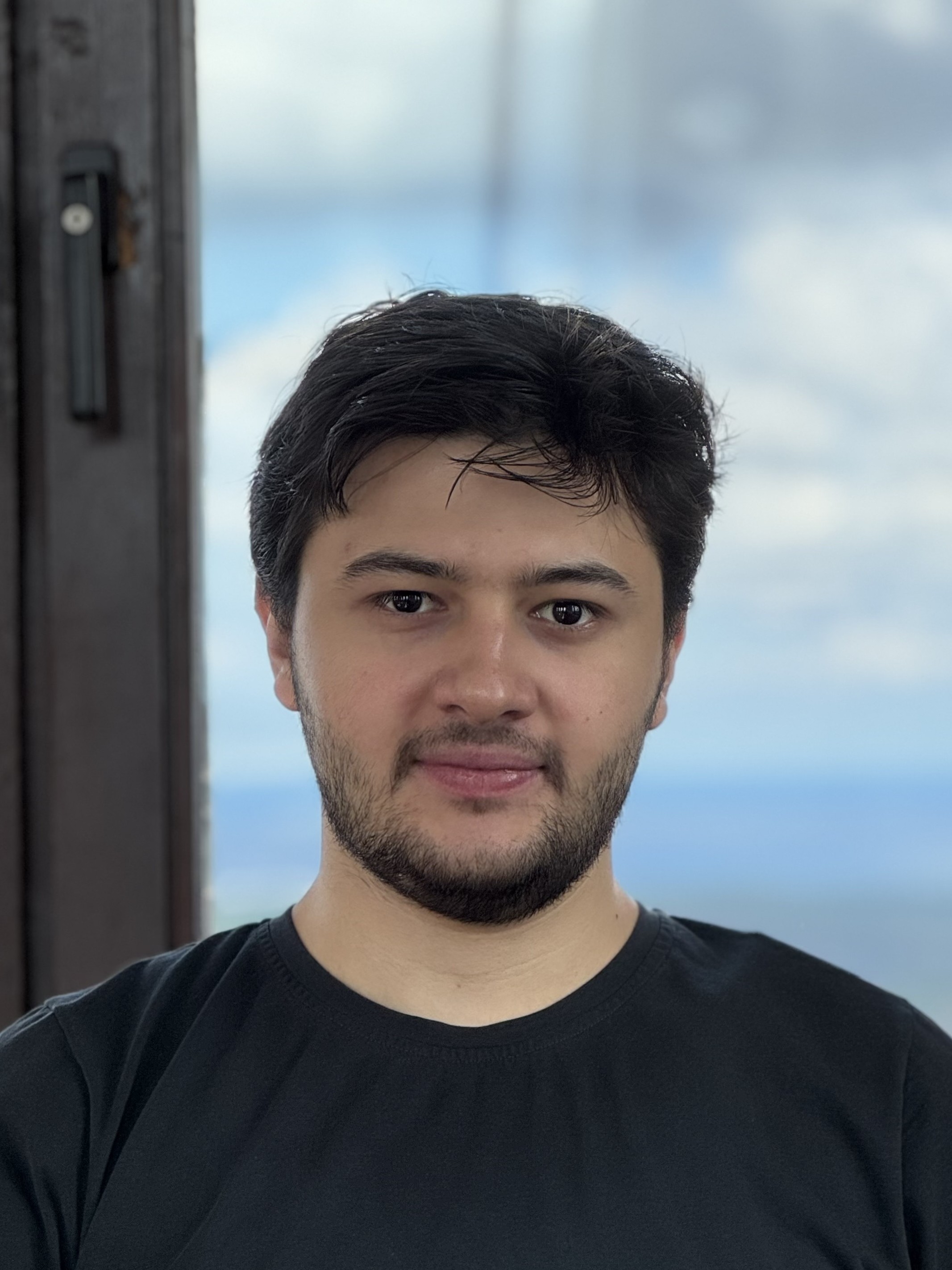}}]{Fatih Merdan}
completed his high school education at Kırıkkale Science High School. He received his B.Sc. degree in Electrical and Electronics Engineering from Middle East Technical University. He is currently pursuing his M.Sc. degree in Electrical and Electronics Engineering under the supervision of Prof. Akan at Koç University, Istanbul, Turkey.
\end{IEEEbiography}

\vspace{11pt}

\begin{IEEEbiography}[{\includegraphics[width=1in,height=1.25in,clip,keepaspectratio]{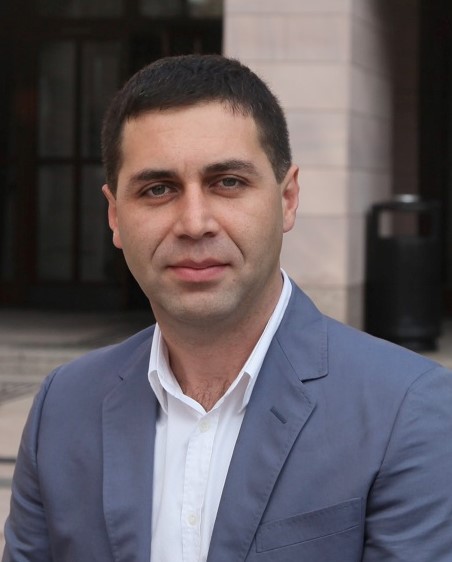}}]{Ozgur B. Akan}
\textbf{(Fellow, IEEE)} received the PhD
from the School of Electrical and Computer Engineering Georgia Institute of Technology Atlanta,
in 2004. He is currently the Head of Internet of
Everything (IoE) Group, with the Department of
Engineering, University of Cambridge, UK and the
Director of Centre for neXt-generation Communications (CXC), Koç University, Turkey. His research
interests include wireless, nano, and molecular communications and Internet of Everything.
\end{IEEEbiography}

\vfill

\end{document}